\documentclass[hidelinks,onefignum,onetabnum,final]{siamart220329}
\usepackage{amsmath,amsfonts,amssymb}
\usepackage{latexsym}
\usepackage{color}
\usepackage{wrapfig}
\usepackage{mathtools} 

\let\eps\varepsilon
\newcommand{\R}{\mathbb R}

\newcommand{\cA}{\mathcal A}

\newcommand{\bC}{\mathbf C}

\newcommand{\bP}{\mathbf P}

\newcommand{\bV}{\mathbf V}

\newcommand{\bg}{\mathbf g}

\newcommand{\be}{\mathbf e}

\newcommand{\bu}{\mathbf u}

\newcommand{\bv}{\mathbf v}

\newcommand{\bw}{\mathbf w}

\newcommand{\bx}{\mathbf x}

\newcommand{\bchi}{{\boldsymbol{\chi}}}

\newcommand{\T}{\mathcal T}

\newcommand{\Div}{\mbox{div}}

\newcommand{\bsigma}{\boldsymbol{\sigma}}

\newcommand{\bPhi}{\boldsymbol{\Phi}}

\newcommand{\rC}{\mathrm C}

\newcommand{\rS}{\mathrm S}
\newcommand{\rU}{\mathrm U}
\newcommand{\rV}{\mathrm V}
\newcommand{\rW}{\mathrm W}

\newcommand{\ru}{\mathrm u}

\newtheorem{remark}{Remark}

\begin{document}
\title{Approximating a  branch  of solutions to the Navier--Stokes equations by reduced-order modeling}
\author{
	Maxim A. Olshanskii\thanks{Department of Mathematics, University of Houston, 
		Houston, Texas 77204 (maolshanskiy@uh.edu).}
		\and
		Leo G. Rebholz\thanks{School of Mathematical and Statistical Sciences, Clemson University, 
		Clemson SC 29634 (rebholz@clemson.edu).}
	}
\maketitle

\begin{abstract}
This paper extends a low-rank tensor decomposition (LRTD) reduced order model (ROM) methodology   to simulate
viscous flows and in particular to predict a smooth branch of solutions for the incompressible Navier-Stokes equations. Additionally, it enhances the LRTD-ROM methodology by introducing a non-interpolatory variant, which demonstrates improved accuracy compared to the interpolatory method utilized in previous LRTD-ROM studies.  After presenting the interpolatory and non-interpolatory LRTD-ROM, we demonstrate that with snapshots from a few different viscosities, the proposed method is able to accurately predict flow statistics in the Reynolds number range $[25,400]$. This is a significantly wider and higher range than state of the art (and similar size) ROMs built for use on varying Reynolds number have been successful on. The paper also discusses how LRTD may offer new insights into the properties of  parametric solutions.
\end{abstract}


\section{Introduction} 
\label{s:intro} 

We are interested in reduced order modeling of the incompressible Navier-Stokes equations (NSE), which are given by
\begin{equation}\label{NSE}
	\left\{
	\begin{aligned}
		\frac{\partial \bu}{\partial t}+(\bu\cdot \nabla)\bu - \nu \Delta \bu +\nabla p&=0, \\
		\Div \bu&=0,
	\end{aligned}\right.
\end{equation}
in a bounded Lipschitz domain $\Omega$ and for $t\in (0,T)$ with a final time $T>0$, and suitable initial and boundary conditions. Here $\bu$ and $p$ are the unknown  fluid velocity  and pressure, and $\nu>0$ is the kinematic viscosity. We treat  $\nu$ as a positive constant parameter that can take values  from the range $[\nu_{\min},\nu_{\max}]$. The problem addressed in the paper consists of an  effective  prediction of flow statistics for 
the entire range $[\nu_{\min},\nu_{\max}]$, based on the information learned from a set of flow states $\bu(t^n,\nu^k),\, p(t^n,\nu^k)$ (further called \textit{snapshots}) computed  for a finite sample of parameters (training set) $\cA=\{\nu^k\}_{k=1}^K\subset [\nu_{\min},\nu_{\max}]$ at given time instances $\{t^n\}_{n=1}^N\subset (0,T]$.     

Assuming $\bu,p$ depends smoothly on $\nu$, the problem outlined above  can be, of course, addressed by numerically solving  \eqref{NSE} for a sufficiently dense sample $\cA$ and proceeding with interpolation. This strategy, however, may entail 	prohibitive computational costs of solving the full order model multiple times for a large set of parameters and impractical data storage. Furthermore, for long-time simulations, such an interpolation strategy may fail to be sustainable given that solution trajectories may (locally) diverge exponentially  fast (although in 2D the system has a finite dimensional attractor~\cite{constantin1985global} which could be captured by  a ROM). 

For a fast and accurate computations of flow statistics for any $\nu\in [\nu_{\min},\nu_{\max}]$, the present paper considers  a reduced order model (ROM) which uses a low-rank tensor decomposition (LRTD) in the space--time--parameter space as the core dimension reduction technique. As such, LRTD replaces SVD/POD, the standard reduction method in more traditional POD--ROMs. This allows for the recovery of information  about the parameter-dependence of reduced spaces from a smaller set of 
pre-computed snapshots and to exploit this information for building parameter-specific ROMs.  The LRTD--ROM was recently introduced in \cite{mamonov2022interpolatory} and further developed and analyzed in \cite{mamonov2023analysis,mamonov2023tensorial}.  

This is the first time LRTD--ROM is applied to the system \eqref{NSE} and, more generally, to predict the dynamics of a viscous fluid flow. The paper extends the approach  to the parameterized incompressible Navier--Stokes equations.  We introduce a non-interpolatory variant of LRTD--ROM. The method is applied to predict drag and lift coefficients for a 2D flow passing a cylinder at Reynolds numbers $\text{Re}\in [25,400]$. This branch of solutions contains the first bifurcation point at around $\text{Re}=50$~\cite{chen1995bifurcation,williamson1996vortex}, when the steady state flow yields to an unsteady periodic flow.  

Predicting flow dynamics along a parameterized  branch of solutions is a challenging task for traditional ROMs, since building a universal low-dimensional space for a range of parameter(s) may be computationally expensive if possible at all. Recent studies that develop or apply reduced order modeling to parameterized fluid problems include~\cite{karatzas2022reduced,pichi2023artificial,hess2023data,hess2023data,reyes2023reduced}.
In particular, several papers addressed the very problem of applying ROMs to predict a flow passing a 2D cylinder for varying Re number.  
The authors of \cite{gao2022reduced} applied dynamic mode decomposition (DMD) with interpolation between pre-computed  solutions for 16 values of viscosity to predict   flow for $\text{Re}\in[85, 100]$. In \cite{andreuzzi2023dynamic} the DMD with 14 viscosity values in the training set was applied to forecast the flow  around the first bifurcation point $R=50$ (the actual Re numbers are not specified in the paper).
A stabilized POD--ROM was tested in \cite{stabile2018finite} to predict the same flow for $\text{Re}\in[100, 200]$.  In \cite{guo2019data} the same problem of the 2D flow around a circular cylinder for varying Re numbers was approached with a POD--ROM based on greedy sampling of the parameter domain. Such POD--ROM required then offline computations of FOM solutions for 51 values of Re to predict flow statistics for $\text{Re}\in [75,100]$. Compared to these studies, the LRTD--ROM is able to handle significantly larger parameter variations with nearly the same or smaller training sets. For example,  we found  13 values of Re log-uniformly sampled to be sufficient for   LRTD--ROM with reduced dimension of 20 to reasonably predict the same flow statistics for $\text{Re}\in [25,400]$.  This exemplifies  the prediction capability of LRTD based projection ROM for fluid problems.

The remainder of the paper is organized as follows. Section~\ref{s:FOM} describes the FOM, which is a second-order in time Scott-Vogelius finite element method on a sequence of barycenter refined  triangulations.  Section~\ref{s:ROM} introduces the reduced order model. In section~\ref{s:num} the model is applied to predict the 2D flow along a smooth branch  of solutions.

\section{Full order model} \label{s:FOM}

To define a full order model for our problem of interest,  we consider a conforming finite element  Galerkin method: Denote by $\bV_h\subset H^1(\Omega)^d$ and
$Q_h\subset L^2_0(\Omega)$ velocity and pressure   finite element spaces with respect to a regular triangulation $\T_h$ of $\Omega$.
For $\bV_h$ and $Q_h$ we choose the lowest order Scott-Vogelius finite element pair:
\begin{equation}\label{wbV}
	\begin{split}
		\bV_h&= \{\bv\in  C(\Omega)^2:\,\bv\in\left[\mathbb{P}_{2}(T)\right]^2~\forall\,T\in\T_h\},\\
		Q_h&= \{q\in L^2(\Omega):\,q\in\mathbb{P}_{1}(T)~~~~\forall\,T\in\T_h\}.
	\end{split}
\end{equation}
The lowest order Scott-Vogelius (SV) element is known \cite{arnold1992quadratic} to be LBB stable in 2D on barycenter refined meshes (also sometimes referred to as Alfeld split meshes). Hence we consider $\T_h$ such that it is obtained by one step of barycenter refinement applied to a coarser triangulation.  Since $\Div(\bV_h)\subseteq Q_h$, it is an example of a stable element which enforces the divergence free constraint for the finite element velocity  pointwise.

Denote by $I_h(\cdot)$ any suitable interpolation operator of velocity boundary values.
We use $(f,g):=\int_\Omega f\cdot g\,dx$ notation for both scalar and vector functions $f,g$. 
We also adopt the notation $\bu_h^n$ and $p_h^n$ for the finite element approximations of velocity and pressure at time $t_n=n\Delta t$, with $\Delta t=T/N$ and $n=0,1,2,\dots,N$.

The second order in time FE Galerkin formulation of \eqref{NSE} with $\bu={\bf g}$ on $\partial \Omega$ reads: Find $\bu_h^n\in\bV_h$, $\bu_h^n=I_h(\bg(t_n))$ on $\partial\Omega$  and $ p_h^n\in Q_h\cap L^2_0(\Omega)$,  for $n=1,2,\dots,N$, such that
satisfying 
\begin{multline}\label{FOM}
		\Big(\frac{3\bu_h^{n}-4\bu_h^{n-1}+\bu_h^{n-2}}{2\Delta t},\bv_h\Big)+((2\bu_h^{n-1}-\bu_h^{n-2})\cdot\nabla \bu_h^n,\bv_h)\\
		+ \nu(\nabla\bu_h^n,\nabla\bv_h)-(p_h^n,\Div\bv_h)+(\Div \bu_h^n,q_h) =0, 
\end{multline}
for all $\bv_h\in\bV_h$, s.t. $\bv_h={\bf 0}$ on $\partial\Omega$, $q_h\in Q_h$,
and $\bu_h^0=\bu(0)$. The first step for $n=1$ is done by the first order implicit Euler method.

The stability and convergence of the method can be analyzed following textbook arguments (e.g.~\cite{girault2012finite,ern2004theory}), implying the estimate
\begin{equation}\label{errEst}
\max_{n=1,\dots,N}\|\bu^n_h-\bu(t^n)\|^2_{L^2(\Omega)}+ \Delta t\nu \sum_{n=1}^N\|\nabla(\bu^n_h-\bu(t^n))\|^2_{L^2(\Omega)} \le C(\bu,p,\nu)(|\Delta t|^4+h^4),
\end{equation}
where $h=\max_{T\in\T_h}\text{diam}(T)$, and $C(\bu,p,\nu)$ is independent on the mesh parameters but depends on the {regularity (smoothness) of $\bu$ and $p$.}
Under extra regularity assumptions, the optimal order velocity and pressure estimates follow in the $L^\infty(L^2)$-norms~\cite{heywood1990finite}:
\begin{equation}\label{errEst2}
	\max_{n=1,\dots,N}(\|\bu^n_h-\bu(t^n)\|_{L^2(\Omega)}+h\|p^n_h-p(t^n)\|_{L^2(\Omega)}) \le C(\bu,p,\nu)(|\Delta t|^2+h^3).
\end{equation}

\section{Reduced order model} \label{s:ROM}

The LRTD--ROM is a projection based ROM, where the  solution  is sought in a parameter dependent low dimensional space. Since the divergence-free finite elements are used for the FOM model, the  low dimensional ROM space  is a velocity space   $\bV^\ell(\nu)\subset \bV_h$, $\mbox{dim}(\bV^\ell(\nu))=\ell\ll M$,  such that 
\begin{equation}\label{divFree}
	\Div\bv_\ell=0\quad\text{for all}~\bv_\ell\in\bV^\ell(\nu).
\end{equation}
Thanks to  \eqref{divFree}, the pressure does not enter the projected equations and the reduced order model reads:
Find $\bu_\ell^n\in\bV_\ell(\nu)$,   for $n=1,2,\dots,N$, such that
satisfy 
\begin{equation}\label{ROM}
	\Big(\frac{3\bu_\ell^{n}-4\bu_\ell^{n-1}+\bu_\ell^{n-2}}{2\Delta t},\bv_\ell\Big)+((2\bu_\ell^{n-1}-\bu_\ell^{n-2})\cdot\nabla \bu_\ell^n,\bv_\ell)
	+ \nu(\nabla\bu_\ell^n,\nabla\bv_\ell) =0, 
\end{equation}
for all $\bv_\ell\in\bV_\ell(\nu)$,
and $\bu_\ell^0=\bP_\ell\bu(0)$, where $\bP_\ell$ is a projector into $\bV_\ell(\nu)$. Similar to the FOM, the implicit Euler method is used for $n=1$.
Once the velocities $\bu_\ell^{n}$ are known, the corresponding pressure functions $p_\ell^n\in Q_h$ can be recovered by a straightforward post-processing step, see, e.g. ~\cite{caiazzo2014numerical}.

The critical part of the ROM is the design of a parameter-specific low-dimensional space $\bV_\ell(\nu)$. 
This is done within a framework of a LRTD--ROM (sometimes referred to as Tensor ROM or TROM), which replaces the  matrix SVD -- a traditional dimension reduction technique -- by   a low-rank tensor decomposition. 
The application of tensor technique is motivated by a natural space--time--parameters structure of the system. 
This opens up possibilities for the fast (online) finding of an efficient low-dimensional  $\nu$-specific ROM space for arbitrary incoming viscosity parameter $\nu$. 
The resulting LRTD--ROM consists of \eqref{ROM}, offline part of applying LRTD, and online part with some fast linear algebra to determine  $\bV_\ell(\nu)$.  Further details of  LRTD--ROM are provided next. 

\subsection{LRTD--ROM}  
Similar to the conventional POD, on an offline stage  a representative collection 
of flow velocity states,  referred to as snapshots, is computed at times $t_j$ and for  pre-selected values of the viscosity  parameter:
\begin{equation*}
	\bu_h(t_j,\nu_k) \in \bV_h, 
	\quad j = 1,\ldots, N, \quad k = 1,\ldots,K.
\end{equation*}
Here $\bu_h$ are  solutions of the full order model~\eqref{FOM}  for a set  of $K$ viscosity parameters $\nu_k\in[\nu_{\min},\nu_{\max}]$.

A standard POD dimension reduction consists then in   finding a subspace $\bV^{\rm pod}_\ell\subset \bV_h$ that approximates the space spanned by all observed snapshots in the best possible way (subject to the choice of the norm). 
This way, the POD reduced order space captures \emph{cumulative} information regarding the snapshots' dependence on the viscosity parameter. Lacking parameter specificity, $\bV^{\rm pod}_\ell$ and so POD--ROM may lack robustness for parameter values outside the sampling set and may necessitate $\ell$ and $K$ to be large to accurately represent the whole branch of solutions.  This limitation motivates the application of a tensor technique based on low-rank tensor decomposition to preserve information about parameter dependence in reduced-order spaces.

Denote by  the upright symbol $\ru^j(\nu)\in\mathbb{R}^{M}$ the vector of representation for $\bu_h(t_j,\nu)$ in the nodal basis.
Recalling that the POD basis can be defined from the  low-rank approximation (given by a truncated SVD) of the snapshot matrix, one can interpret  LRTD as a multi-linear extension of POD:  Instead of arranging snapshots in a matrix $\Phi_{\rm pod}$,  one seeks to exploit the  tensor structure of the snapshots domain and to utilize the LRTD instead of the matrix SVD for solving a tensor analogue of the low-rank approximation problem.


For LRTD--ROM, the  coefficient vectors of velocity snapshots are  organized in the \emph{multi-dimensional} array 
\begin{equation}
(\bPhi)_{:,k,j} = \ru^j(\nu_k), 
\label{eqn:snapmulti}
\end{equation}
which is a 3D tensor of  size $M\times K\times N$. The first and the last  indices of $\bPhi$ correspond to the spatial and temporal  dimensions, respectively.

Unfolding of  $\bPhi$ along its first mode into a  $M\times NK$ matrix and proceeding with its truncated SVD constitutes the traditional POD approach.
In the tensor ROM the truncated SVD of the unfolded matrix  is replaced with a truncated LRTD of $\bPhi$.

Although the concept of tensor rank is somewhat ambiguous, there is an extensive literature addressing the issue of defining tensor rank(s) and LRTD; see e.g.~\cite{hackbusch2012tensor}. 
In \cite{mamonov2022interpolatory,mamonov2023tensorial}, the tensor ROM has been introduced for three common rank-revealing tensor formats: canonical polyadic, Tucker, and tensor train. The LRTD in any of these  formats can be seen as an extension of the SVD to multi-dimensional arrays.  While each format has its  distinct numerical and compression properties and would be suitable, we use Tucker  for the purpose of this paper.  

We note that the  LRTD approach is effectively applicable for multi-parameter problems. In case of a higher parameter space dimension one may prefer a hierarchical  Tucker format~\cite{hackbusch2009new} such as tensor train to avoid exponential growth of LRTD complexity with respect to the parameter space dimension.

In the Tucker format ~\cite{tucker1966some,ReviewTensor}
one represents  $\bPhi$ by the following sum of direct products of three vectors:
\begin{equation}
	\label{eqn:TDv}
	\bPhi \approx \widetilde{\bPhi} = 
	\sum_{m = 1}^{\widetilde{M}}
	\sum_{k = 1}^{\widetilde{K}}
	\sum_{n = 1}^{\widetilde{N}}
	(\bC)_{m,k,n} \bw^m \otimes \bsigma^{k} \otimes \bv^n,
\end{equation}
with $\bw^m \in \R^M$, $\bsigma^{k} \in \R^{K}$, and $\bv^n \in \R^N$. Here $\otimes$ denotes the outer vector product.
The numbers $\widetilde{M}$, $\widetilde{K}$, and $\widetilde{N}$ 
are referred to as Tucker ranks of $\widetilde{\bPhi}$. The Tucker format delivers an efficient compression 
of the snapshot tensor, provided the size of the \emph{core tensor} 
$\bC$ is (much) smaller than the size of $\bPhi$, i.e.,  $\widetilde{M} \ll M$, $\widetilde{K} \ll K$,  and $\widetilde{N} \ll N$. 

Denote by  $\|\bPhi\|_F$ the tensor Frobenius norm, which is  the  square root of the sum of the squares of all  entries of $\bPhi$.
Finding the best approximation of a tensor in the Frobenius norm by a fixed-ranks Tucker tensor  is a well-posed problem with a constructive algorithm  known to deliver quasi-optimal solutions~\cite{ReviewTensor}.
Furthermore, using this algorithm, which is  
based on the truncated SVD for a sequence of unfolding matrices, one finds $\widetilde{\bPhi}$ in the Tucker format that satisfies
\begin{equation}
	\label{eqn:TensApproxF}
	\big\| \widetilde{\bPhi} - \bPhi \big\|_F \le  \widetilde\eps \big\|\bPhi \big\|_F
\end{equation}
for a given $\widetilde\eps > 0$ and the sets $\{\bw^m\}$, $\{\bsigma^k\}$, $\{\bv^n\}$ are orthogonal. Corresponding Tucker ranks are then recovered in the course of factorization.  The resulting  decomposition  for $\widetilde\eps=0$ is also known as Higher Order SVD (HOSVD) of $\bPhi$~\cite{de2000multilinear}.

%

For arbitrary but fixed $\nu\in[\nu_{\min},\nu_{\max}]$, one can `extract'  from $\widetilde{\bPhi}$ specific (local) information for building $\bV_\ell(\nu)$.
We consider two approaches herein: The first one adopts interpolation between available snapshots but is done directly in the low-rank format, while another avoids the interpolation step.

\subsubsection{Interpolatory LRTD--ROM}
To  formulate interpolatory LRTD--ROM, we need several further notations.
We assume an interpolation procedure 
\begin{equation}
	\label{eqn:bea}
	\bchi \,:\, [\nu_{\min},\nu_{\max}] \to \mathbb{R}^{K}
\end{equation}
such that for  any smooth function $g:\,[\nu_{\min},\nu_{\max}]\to \R$,
$
I(g):= \sum_{k=1}^{K} \bchi (\nu )_kg(\nu_{k})
$
defines an interpolant for $g$.
Our choice is the Lagrange interpolation of order $p$: 
\begin{equation}
	\label{eqn:lagrange}
	\bchi (\nu)_k = 
	\begin{cases} 
		\prod\limits_{\substack{m = 1, \\ m \neq k}}^{p}(\nu_{i_m}-\nu) \Big/ 
		\prod\limits_{\substack{m = 1, \\ m \neq k}}^{p}(\nu_{i_m}-\nu_k), 
		& \text{if } k = i_k \in \{i_1,\ldots,i_p\}, \\
		\qquad\qquad\qquad\qquad\qquad\qquad\quad 0, & \text{otherwise}, \end{cases}
\end{equation}
where  $\nu_{i_1}, \ldots, \nu_{i_p}\in [\nu_{\min},\nu_{\max}]$ 
are the $p$ closest to $\nu$  viscosity values from the training set.

The $\nu$-specific \emph{local} reduced  space $V^\ell(\nu)$  is the space spanned by the first  $\ell$ left singular vectors of the  matrix $\widetilde{\Phi} (\nu)$, defined through the in-tensor interpolation procedure for $\widetilde{\bPhi}$:
\begin{equation}
	\label{eqn:extractbt}
	\widetilde{\Phi} (\nu) = \widetilde{\bPhi} 
	\times_2 \bchi(\nu) 
	\in \R^{M \times N},
\end{equation}
where $\times_2$ denotes the tensor-vector multiplication along the second mode.

Consider a nodal basis denoted as ${\xi_h^j}$ in the finite element velocity space $\bV_h = \text{span}\{\xi_h^1, \dots, \xi_h^M\}$.
The corresponding finite element LRTD--ROM  space is then  
\begin{equation}
	\label{Vromh}
	\begin{split}
		\bV^\ell(\nu)&=\{\bv_h\in \bV_h\,:\, \bv_h=\sum_{i=1}^M\xi^i_h(\bx)v_i,~~\text{for}~(v_1,\dots,v_M)^T\in V^\ell(\nu)\},\\
		&\text{where}\quad 
		V^\ell(\nu)=\mbox{range}(\rS(\nu)(1:\ell)),~\text{for}~ \{\rS(\nu),\Sigma(\nu), \rV(\nu)\}=\text{SVD}(\widetilde{\Phi} (\nu)).
	\end{split}
\end{equation}
In section~\ref{s:Impl} we will discuss implementation details omitted here. 

\subsubsection{Non-interpolatory LRTD--ROM}
In  non-interpolatory LRTD--ROM, the basis of the local ROM space is built as an optimal $\ell$-dimensional space approximating the space spanned by snapshots corresponding to several nearest in-sample viscosity values. For this we need the {extraction} 
procedure   
\begin{equation}
	\label{eqn:extractbt2}
	\widetilde{\Phi}_k = \widetilde{\bPhi} 
	\times_2 \be_k 
	\in \R^{M \times N},
\end{equation} 
so that $\widetilde{\Phi}_k$ is the $k$-th space-time slice of $\widetilde{\bPhi}$.

As in the interpolatory LRTD--ROM, let  $\nu_{i_1}, \ldots, \nu_{i_p}\in [\nu_{\min},\nu_{\max}]$ be the $p$ closest to $\nu$ sampled viscosity values.
Then the $\nu$-specific \emph{local} reduced space $V^\ell(\nu)$  is the space spanned by the first  $\ell$ left singular vectors the following low-rank matrix $\widetilde{\Phi} (\nu)$:
\begin{equation}
	\label{eqn:extractbt3}
	\widetilde{\Phi} (\nu) = [\widetilde{\Phi}_{i_1},\dots,\widetilde{\Phi}_{i_p}]
	\in \R^{M \times pN}.
\end{equation}
The corresponding finite element LRTD--ROM  space is defined in the same way as in \eqref{Vromh}. \medskip

A remarkable feature of the LRTD--ROM is that finding the basis of $V^\ell(\nu)$, i.e. finding the left singular vectors of $\widetilde{\Phi} (\nu)$, does not require building or working with the 'large' matrix    $\widetilde{\Phi} (\nu)$.
For any given $\nu\in[\nu_{\max},\nu_{\min}]$ it involves calculations with lower dimension objects only and so it can be effectively done online. This implementation aspect of   LRTD--ROM is recalled below.

\subsubsection{Implementation} \label{s:Impl}
The implementation of the Galerkin LRTD-ROM  follows a two-stage procedure.

\textit{Offline stage.} For a set of sampled viscosity parameters, the snapshot tensor $\bPhi$ is computed and for chosen $\eps>0$ the truncated HOSVD is used to find  $\widetilde{\bPhi}$ satisfying \eqref{eqn:TensApproxF}. This first stage defines the \emph{universal reduced space} $\widetilde{V}$, which is the span of all $\bw$-vectors from the Tucker decomposition \eqref{eqn:TDv}:
\begin{equation}\label{Univ}
	\widetilde{V} = \text{span}\big\{\bw_1,\dots,\bw_{\widetilde{M}}\big\}\subset\R^M.
\end{equation}
Hence the dimension of $\widetilde{V}$ is equal to  the first Tucker rank $\widetilde{M}$ and $\bw_1,\dots,\bw_{\widetilde{M}}$ is an orthonormal basis. 
At this point the system \eqref{FOM} is `projected' into $\widetilde{V}$, i.e. 
the  projected velocity mass, stiffness matrices, initial velocity, and the projected inertia term 
are passed to the online stage.  

\begin{remark}[Nonlinear terms]\rm To handle the inertia term, we benefit from its quadratic non-linearity. More precisely, we compute a sparse 3D array 
	\[
	\mathrm{N}\in\R^{M\times M\times M},\quad \text{with entries}~ \mathrm{N}_{ijk}=(\xi_i\cdot\nabla\xi_j,\xi_k),  
	\]
and project it  into  $\widetilde{V}$  by computing 	
	\[
	\widetilde{\mathrm{N}}=\mathrm{N}\times_1\mathrm{W} \times_2\mathrm{W}\times_3\mathrm{W}, 
\]    
where $\mathrm{W}=[\bw_1,\cdots,\bw_{\widetilde{M}}]\in\R^{M\times\widetilde{M}}$ and
$\times_i$ is now the tensor-matrix product along $i$-th mode. The $ {\widetilde{M}\times\widetilde{M}\times\widetilde{M}}$ array $\widetilde{\mathrm{N}}$ is passed to the online stage.

An alternative, which we do not pursue here, would be the application of a LRTD--DEIM technique~\cite{mamonov2023tensorial} to handle the nonlinear terms. 
\end{remark}

\textit{Online stage.}  The online stage receives the projected matrices, and the 3D array 
$\widetilde{\mathrm{N}}$. From the LRTD \eqref{eqn:TDv} it  receives  the core tensor $\bC$  and the matrix  $\rS = [\bsigma^1, \dots, \bsigma^{\widetilde{K}}]^T$.  

To find $V^\ell(\nu)$ for any $\nu\in [\nu_{\min},\nu_{\max}]$, one first computes a local core matrix
\begin{equation}\label{Ce}
	\rC(\nu) = 
	\left\{
	\begin{split}
	&\bC \times_2 \left(\rS  \bchi(\nu) \right)  \in \R^{\widetilde{M} \times \widetilde{N}} &\text{\footnotesize interpolatory case},\\
	&\left[\bC \times_2 \left(\rS \be_{i_1} \right),\dots, \bC \times_2 \left(\rS \be_{i_p} \right) \right]  \in \R^{\widetilde{M} \times p\widetilde{N}} &\text{\footnotesize non-interpolatory case}. 
	\end{split}
	\right.
\end{equation}
and  its thin SVD, 
$
		\rC(\nu) = \rU_c \Sigma_c \rV_c^T.
$
It can be easily seen that 
	\begin{equation}\label{SVDc}
		\widetilde{\Phi} (\nu) = 
		\left({\rW}\rU_c\right)\Sigma_c \mathrm{Y}^T,
	\end{equation}
with an orthogonal matrix $ \mathrm{Y}$.	Since the local ROM space $V^\ell(\nu)$ is spanned by the first $\ell$ singular vectors of $\widetilde{\bPhi}(\nu)$ and \eqref{SVDc} is the SVD of $\widetilde{\Phi} (\nu)$,   \emph{the coordinates}
of the local reduced basis in the universal basis $\{\bw_i\}_{i=1}^{\widetilde{M}}$ are the first $\ell$ left singular vectors of 
$\rC(\nu)$, i.e. the first $\ell$ columns of $\rU_c$.  The pre-projected initial velocity, mass and stiffness matrices and
 $\widetilde{\mathrm{N}}$ are projected further into  $V^\ell(\nu)$. The projection is done through multiplication with matrix $\rU_c$.   This allows the execution of the proposed ROM \eqref{ROM}. 
 
 If the ROM  needs to be rerun for a different value of $\nu$, only calculations starting with \eqref{Ce} {need to be} redone, without any reference to the offline stage data.  
 \medskip

\begin{table}[h!]
	\begin{center}
		\begin{tabular}{r@{\hspace{2mm}}|c@{\hspace{1mm}}c@{\hspace{0mm}}c@{\hspace{3mm}}c@{\hspace{3mm}}c@{}}
			&\multicolumn{3}{c}{Offline part} 	& & \hskip-7ex{Online part}\\[-2.5ex]\\
			&\multicolumn{3}{c}{$\overbrace{\mbox{\hspace{28ex}}}$} &\multicolumn{2}{c}{\hskip-4ex$\overbrace{\mbox{\hspace{21ex}}}$}  \\
		& & \hskip-3ex$-\text{\scriptsize LRTD}\hskip-0.6ex\rightarrow$\hskip-4ex    & & & \hskip-20ex$-\text{\scriptsize LRLA}\hskip-0.6ex\rightarrow$  \\
Spaces		& $\bV_h$             & $\supset$ & $\widetilde{\bV}_h$                         & $\supset$ &\hskip-4ex $\bV^{\ell}(\nu)$\\[-0.5ex]
	&	$\shortparallel$  &   &\hskip-3ex  $\shortparallel$ &&\hskip-7ex $\shortparallel$\\[-0.5ex]
		& $\mbox{span}\{\xi_h^i\}_{i=1}^M$&     &	$\mbox{span}\{w_h^i\}_{i=1}^{\widetilde M}$ &      & 	$\mbox{span}\{u_h^i(\nu)\}_{i=1}^\ell$\\[1.2ex]
		\hline \\[-1ex]
Matrices		& {\small FOM }&   &\small Projected&  &\small  Double-projected\\
		&\small  matrices  &         &\small  matrices&&\small  matrices\\[1.5ex]
		\end{tabular}
	\end{center}
\caption{\label{tab:1} Data structure of the LRTD--ROM.  LRLA stands for ``low-rank linear algebra'', which means that all calculations are done with low-dimension objects.}
\end{table} 

We summarize the structure of LRTD-ROM in Table~\ref{tab:1}. The intermediate finite element space  $ \widetilde{\bV}_h$ is the finite element counterpart of the universal space $\widetilde{V}$ from \eqref{Univ}. The basis $\{w_h^i\}_{i=1}^{\widetilde M}$ of $ \widetilde{\bV}_h$ is given in terms of its coordinates in the nodal basis $\{\xi_h^i\}_{i=1}^M$.  In turn, the basis $\{u_h^i(\nu)\}_{i=1}^\ell$ of the $\nu$-specific  local space $\bV^{\ell}(\nu)$ is given by its coordinates in $\{w_h^i\}_{i=1}^{\widetilde M}$. Hence, FE matrices are first projected on  $ \widetilde{\bV}_h$ during the offline phase. They are stored online and double-projected
 for any incoming $\nu$ before executing the ROM~\eqref{ROM}. In general, it holds $\mbox{dim}(\bV_h)\gg\mbox{dim}(\widetilde{\bV}_h) \gg \mbox{dim}(\bV^{\ell}(\nu))$, e.g. in the example from the next section we have $\mbox{dim}(\bV_h)=121,064$, $\mbox{dim}(\widetilde{\bV}_h)=404$, and $\mbox{dim}(\bV^{\ell}(\nu))=20$.

\section{Numerical tests}
\label{s:num} 
{ 

We now test the proposed LRTD--ROM on a benchmark test for incompressible Navier-Stokes flow.  After describing the test problem setup, FOM and ROM construction details, we test the proposed ROM's accuracy in predicting a branch of solutions for the Navier-Stokes equations for $Re$ in [25,400] using snapshots from solutions using 13 and 25 different viscosities.

 \subsection{Test problem description}

The test problem we consider is 2D channel flow past a cylinder \cite{ST96}.  The domain is  [0, 2.2]$\times$[0, 0.41], which represents a rectangular channel, and with a cylinder centered at $(0.2,0.2)$ having radius $0.05,$ see Figure \ref{cylinderdomain}.
 
 	\begin{figure}[!ht]
		\centering
		\includegraphics[scale = .8]{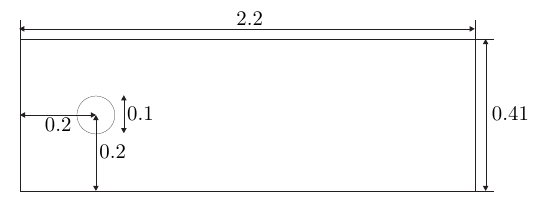}
		\caption{\label{cylinderdomain} Shown above is the domain for the flow past a cylinder test problem.}
	\end{figure}
	
 There is no external forcing for this test, no-slip boundary conditions are enforced on the walls and cylinder, and an inflow/outflow profile 
 	\begin{align*}
	u_1(0,y,t) & = u_1(2.2,y,t) = \frac{6}{0.41^2}y(0.41-y),
	\\u_2(0,y,t) & = u_2(2.2,y,t) = 0
	\end{align*}
	is enforced as a Dirichlet boundary condition.  Of interest for comparisons and accuracy testing are the predicted lift and drag, and for these quantities we use the definitions
	\begin{align*}
	c_d(t) &= 20\int_S \left( \nu \frac{\partial u_{t_S}(t)}{\partial n}n_y - p(t)n_x \right)dS,
	\\c_l(t) &= 20 \int_S \left( \nu \frac{\partial u_{t_S}(t)}{\partial n}n_x - p(t)n_y \right)dS,
	\end{align*}
	where $u_{t_S}$ is the tangential velocity, $S$ the cylinder, and $n = \langle n_x, n_y\rangle$ is the outward unit normal vector. For the calculations, we used the global integral formulation from \cite{J04}.

\subsection{Full order model simulations}

 To study the performance of the LRTD--ROM with respect to the spatial mesh refinement,  we consider three regular  triangulations  $\T_h$ of $\Omega$. The finest triangulation consists of 62,805 triangles, while the inter-medium and the coarsest meshes have 30,078 and 8,658 triangles;  the coarsest mesh is illustrated in Figure~\ref{coarsemesh}.  We note the triangulations are constructed by first creating a Delaunay triangulation followed by a barycenter refinement (Alfeld split).  All FOM simulations used the scheme \eqref{FOM} with time step $\Delta t=0.002$, and lowest order Scott-Vogelius elements as described in section 2.  With this choice of elements, the three meshes provided 252,306, 121,064 and 35,020 total spatial degrees of freedom(dof).  For a given viscosity $\nu$, the corresponding Stokes solution was found with this element choice and mesh to generate the initial condition.

	\begin{figure}[!ht]
		\centering
		\includegraphics[scale = .18]{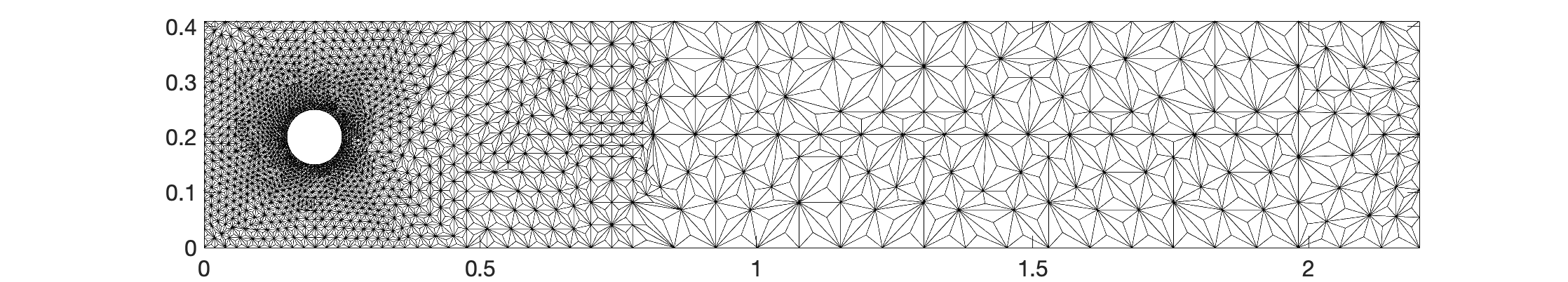}
		\caption{\label{coarsemesh} Shown above is the coarsest mesh used for the flow past a cylinder test simulations.}
	\end{figure}

The viscosity parameter sampling  set consists of $K$ viscosity values log-uniformly distributed over the interval [$2.5 \cdot 10^{-4}, 4 \cdot 10^{-3}$], which corresponds to $25\le Re\le 400$. $K=25$ was the maximum value we used for training the ROM, and results are shown below for varying $K$.

 All FOM simulations were run for $t\in[0,6]$, and by $t=5$ the von Karman vortex street was fully develop behind the cylinder for $Re\gtrapprox50$.  For $Re<50$, the flows had reached a steady state by $t=5$. For each $\nu$ from the sampling set, 251 velocity snapshots were collected for $t\in[5,6]$ in uniformly distributed time instances. This resulted in a $M\times K \times N$ snapshot tensor $\bPhi$, with $M$=dof for each mesh, $K$ different viscosities, and $N=251$.   We note that the Stokes extension of the boundary conditions (which is very close to the snapshot average but preserves an energy balance \cite{MRXT17}) was subtracted from each snapshot used to build $\bPhi$.
  
 \subsection{ROM basis and construction}

Table~\ref{tab:ranks} shows the dependence of the tensor $\widetilde{\bPhi}$ ranks on the targeted compression accuracy $\eps$ and the FOM spatial resolution. The first rank determines the universal space dimension.
As expected, higher ranks are needed for better accuracy. At the same time the dependence on spatial resolution is marginal.

\begin{table}[h]

	\small
	\begin{center}
		\begin{tabular}{c|c|c|c}
			\hline
			
			& Mesh 1 & Mesh 2  &Mesh 3 \\
			target accuracy / M       &35020 & 121064  &252306\\      
			\hline
			$\eps=10^{-1}$ & [15,12 ,7] &[15,11,7]   & [18,13,8]\\
			$\eps=10^{-2}$ & [74,21,40]  & [78,21,40]  & [89,21,45]\\
			$\eps=10^{-3}$ & [190 ,22 ,80] &[213,22,85]   & [239,22,93]\\
			$\eps=10^{-4}$ & [365,23,113]  & [404,23,124]  & [444,23,135]\\
			\hline
		\end{tabular}
			\caption{ 	\label{tab:ranks} HOSVD ranks of the $\eps$-truncated LRTD for the snapshot tensor.}

	\end{center}
\end{table}

\begin{figure}[h]\caption{\label{fig:sigma} Singular values decay for POD matrix and local LRTD matrix for 10 parameter values. Left and right panels show result for interpolatory and non-interpolatory LRTD--ROMs, respectively.}
	\centering
	\includegraphics[width=0.4\textwidth]{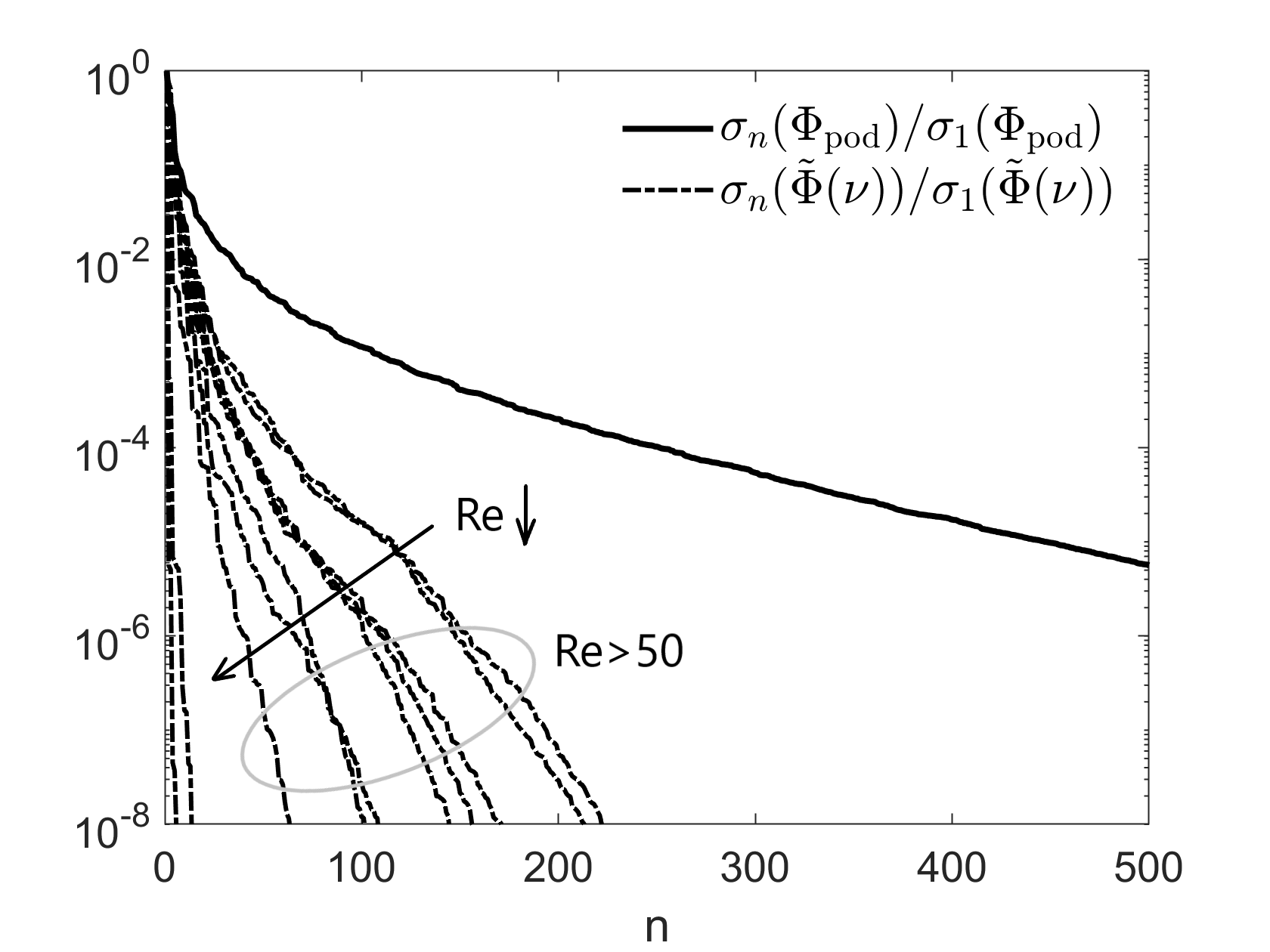}\qquad
	\includegraphics[width=0.4\textwidth]{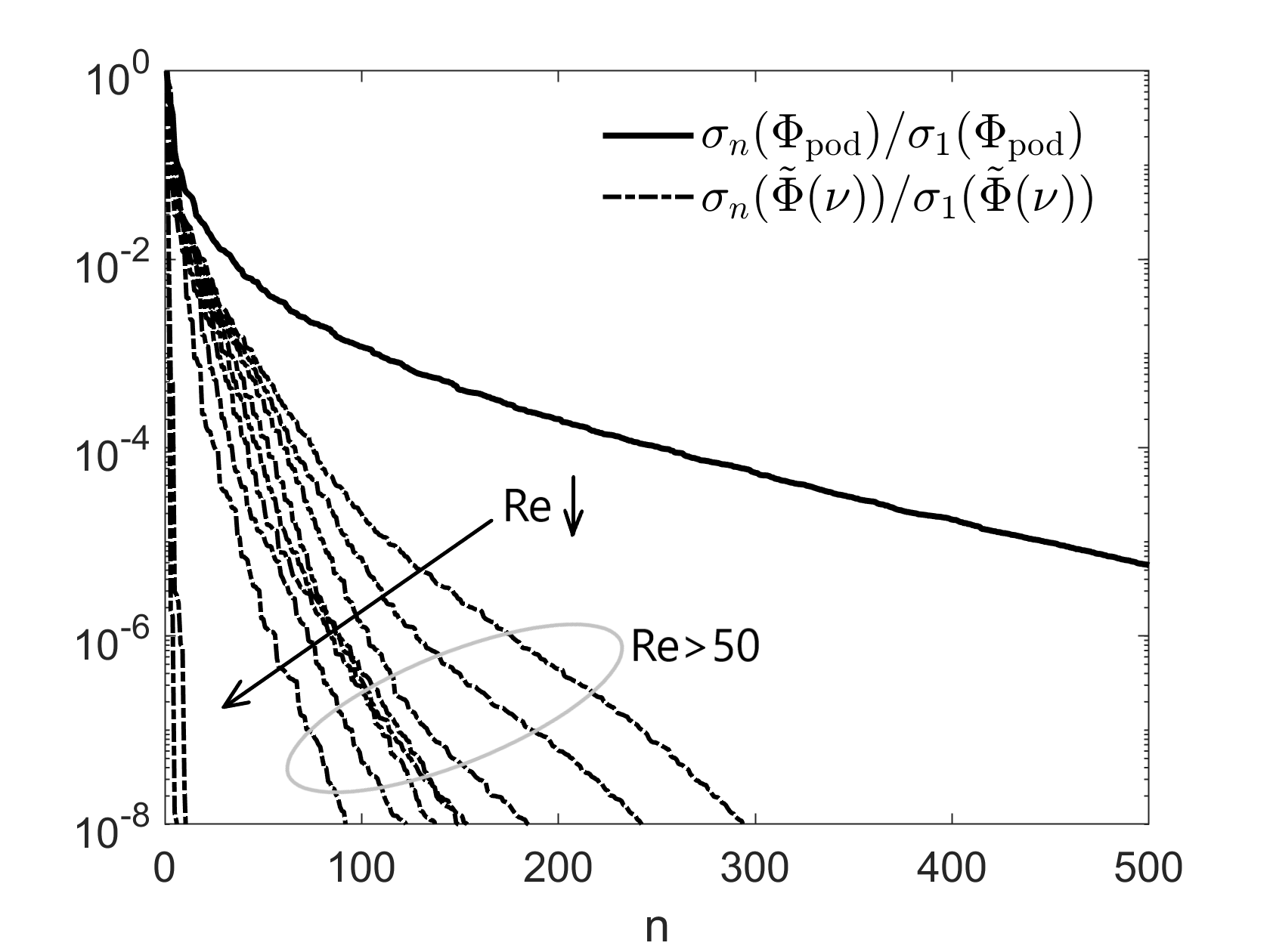}
\end{figure}

Figure~\ref{fig:sigma} illustrates on why finding $\nu$-dependent  local ROM spaces through LRTD is beneficial compared to  employing a POD space, which is universal for all parameter values. The faster decay of singular values $\sigma(\tilde\Phi(\nu))$ allows for attainment of the desired accuracy using lower ROM dimensions compared to the  POD--ROM. At the same time, the decay rate of  $\sigma(\tilde\Phi(\nu))$ depends on $\nu$ with faster decay observed for larger viscosity values, i.e. smaller Reynolds numbers. Unsurprisingly, the snapshots collected for $\text{Re}<50$ show very little variability, indicated by the abrupt decrease of $\sigma_n$ for $n>1$, since the flow approaches an equilibrium state in these cases.    

A plot of the first 7 modes for non-interpolatory LRTD--ROM using $\epsilon=10^{-4}$ with $Re=$110 and 380, and for the full POD constructed with data from tests using all the parameter values, are shown in figure \ref{modes}.  We observe that for the LRTD--ROM cases, the modes quickly go from larger scales in the first few modes to much finer scales by mode 7, whereas for the full POD, the first 7 mode are all still at larger scales.  This is consistent with figure \ref{fig:sigma}, which shows the decay of singular values is much slower, meaning more modes are needed to characterize the space.

\begin{figure}[h]	\centering
	\includegraphics[width = .32\textwidth, height=.13\textwidth,viewport=80 20 580 170, clip]{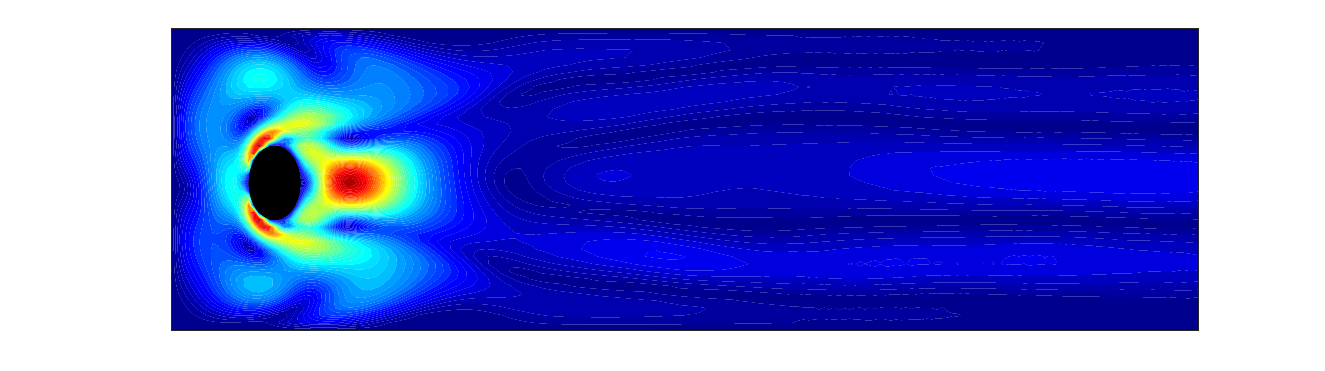}
	\includegraphics[width = .32\textwidth, height=.13\textwidth,viewport=80 20 580 170, clip]{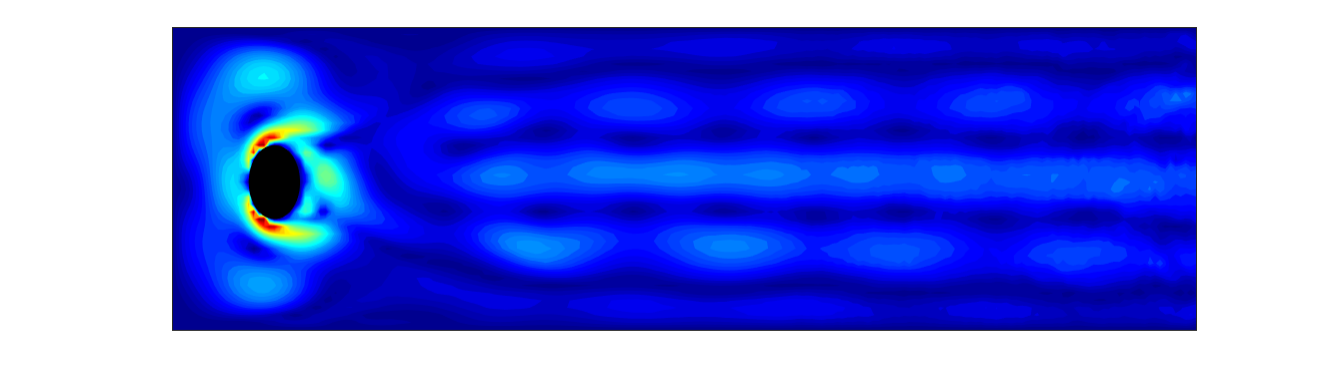}
	\includegraphics[width = .32\textwidth, height=.13\textwidth,viewport=80 20 580 170, clip]{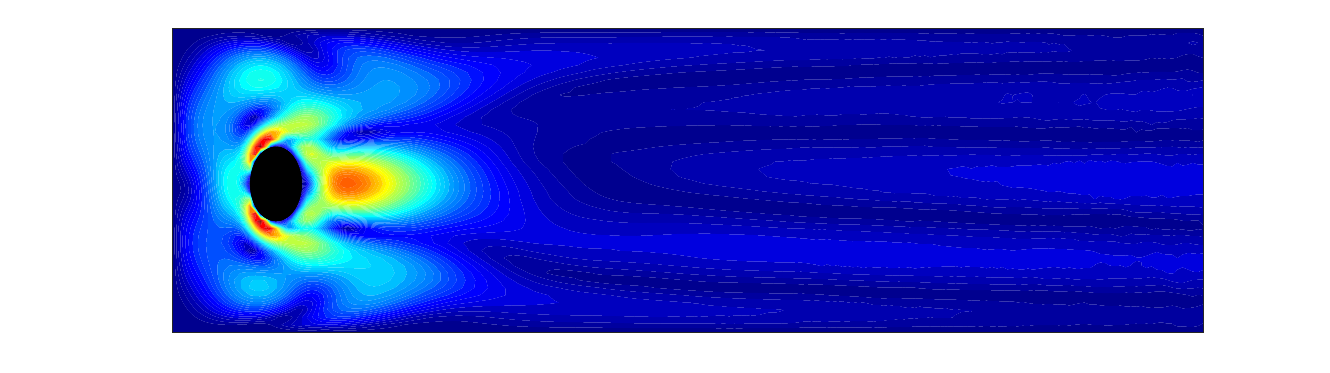}\\
	\includegraphics[width = .32\textwidth, height=.13\textwidth,viewport=80 20 580 170, clip]{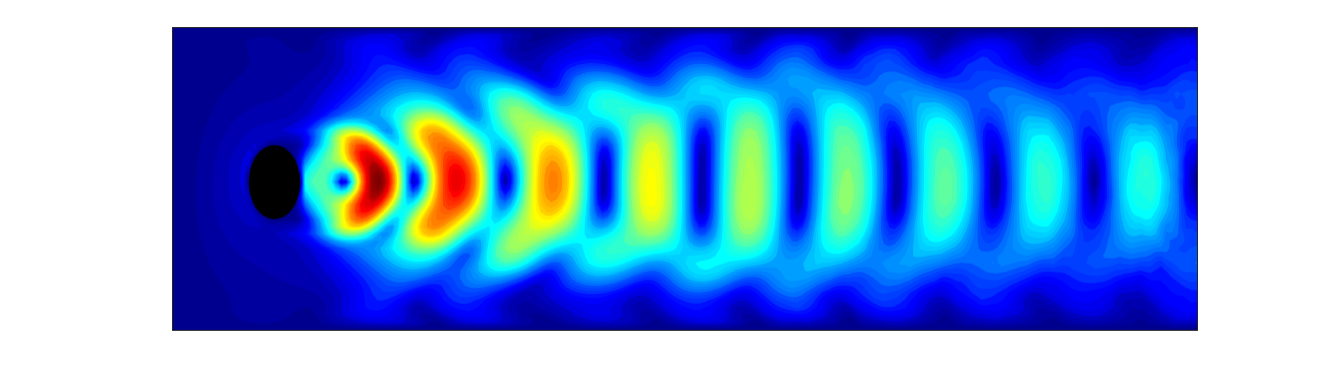}
	\includegraphics[width = .32\textwidth, height=.13\textwidth,viewport=80 20 580 170, clip]{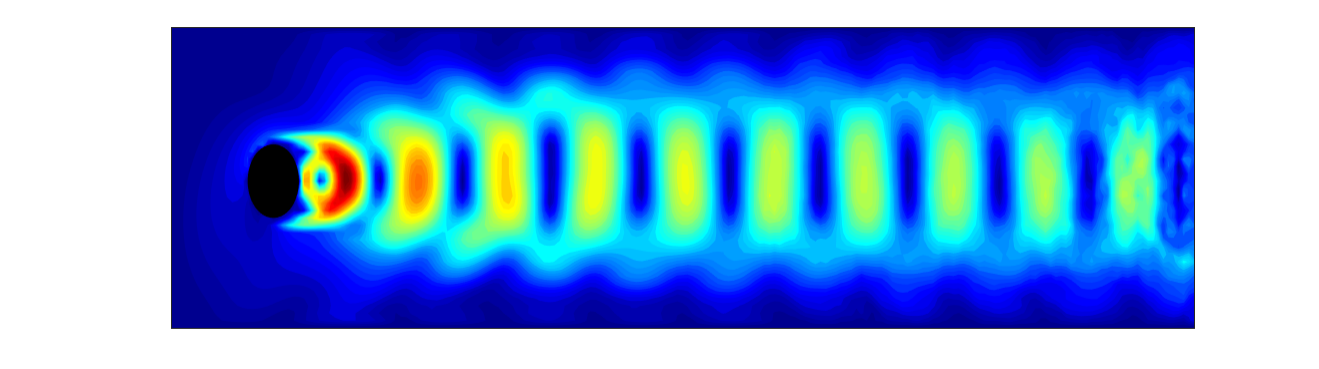}
	\includegraphics[width = .32\textwidth, height=.13\textwidth,viewport=80 20 580 170, clip]{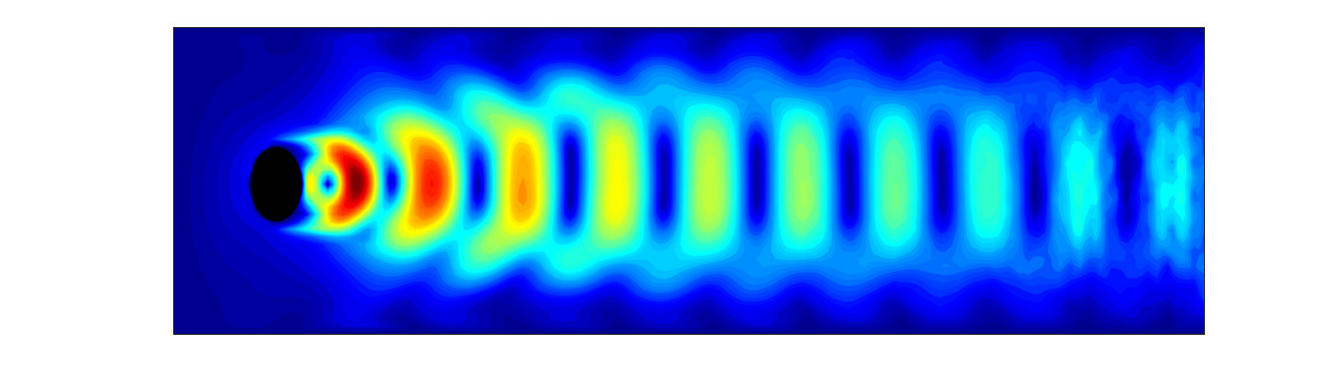}\\
	\includegraphics[width = .32\textwidth, height=.13\textwidth,viewport=80 20 580 170, clip]{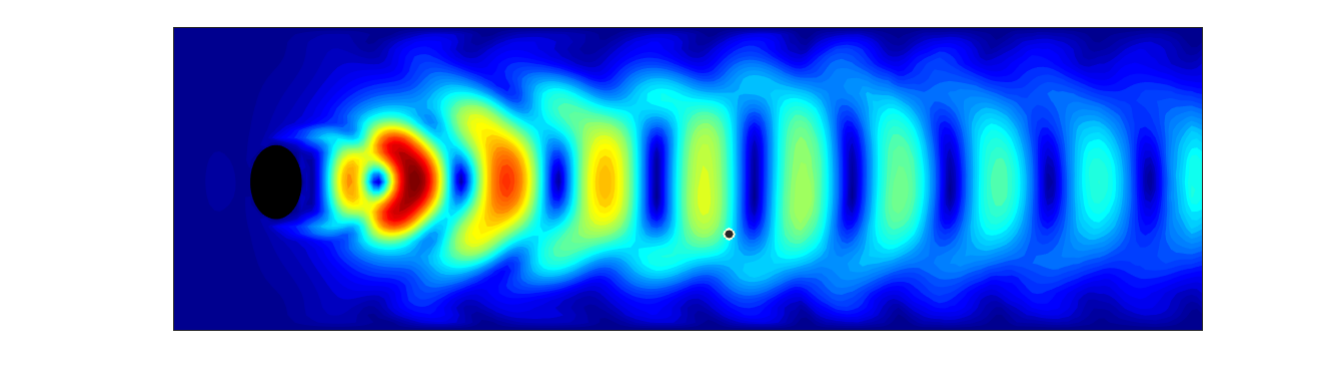}
	\includegraphics[width = .32\textwidth, height=.13\textwidth,viewport=80 20 580 170, clip]{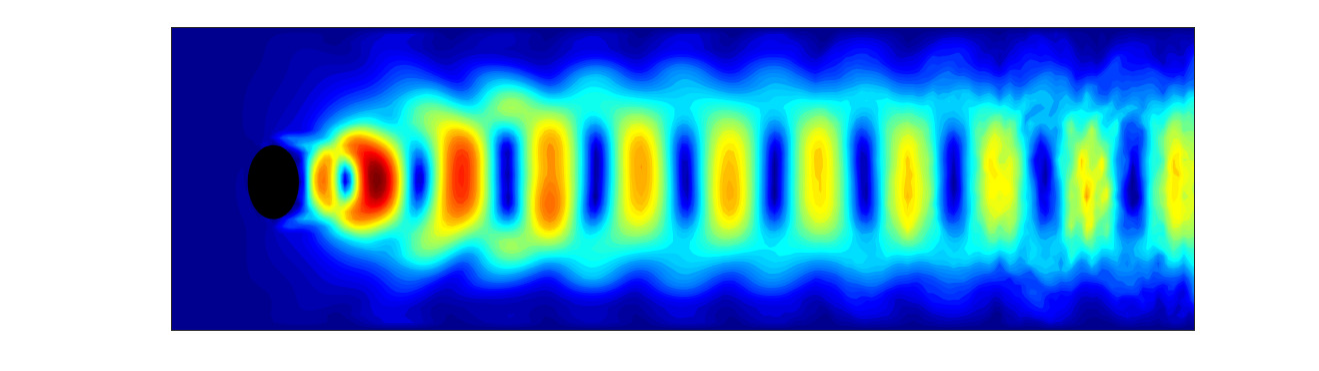}
	\includegraphics[width = .32\textwidth, height=.13\textwidth,viewport=80 20 580 170, clip]{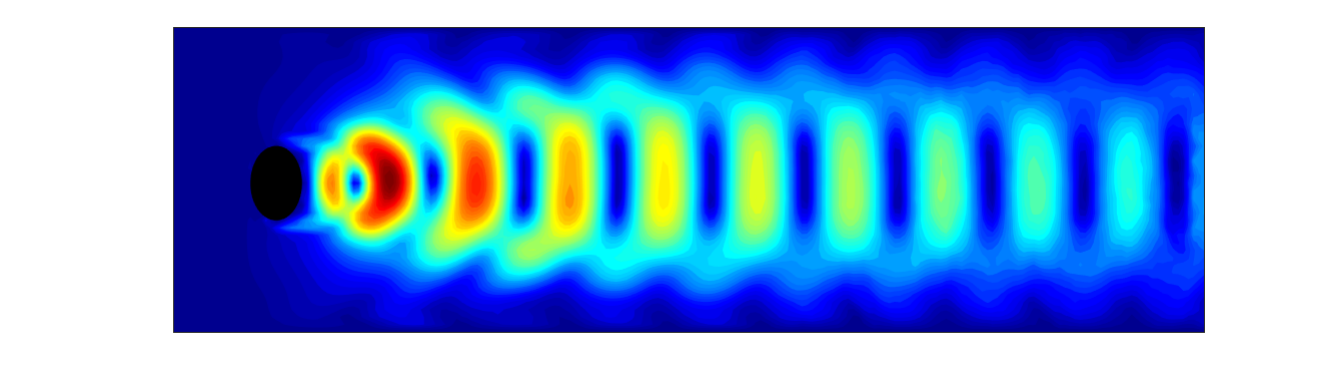}\\
	\includegraphics[width = .32\textwidth, height=.13\textwidth,viewport=80 20 580 170, clip]{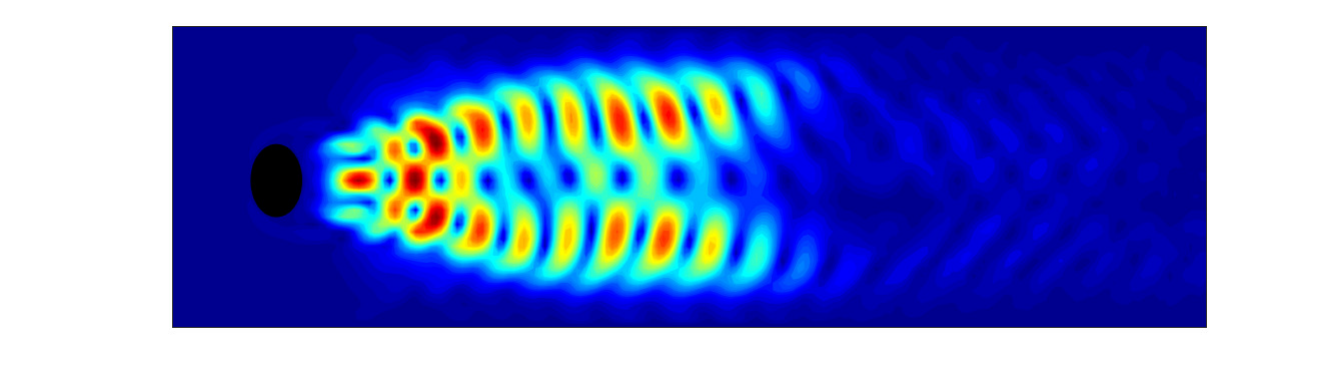}
	\includegraphics[width = .32\textwidth, height=.13\textwidth,viewport=80 20 580 170, clip]{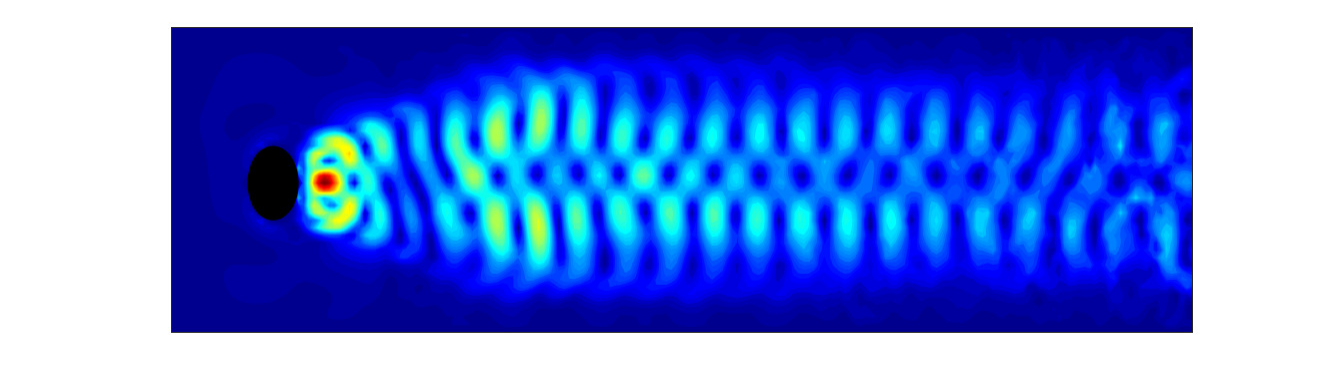}
	\includegraphics[width = .32\textwidth, height=.13\textwidth,viewport=80 20 580 170, clip]{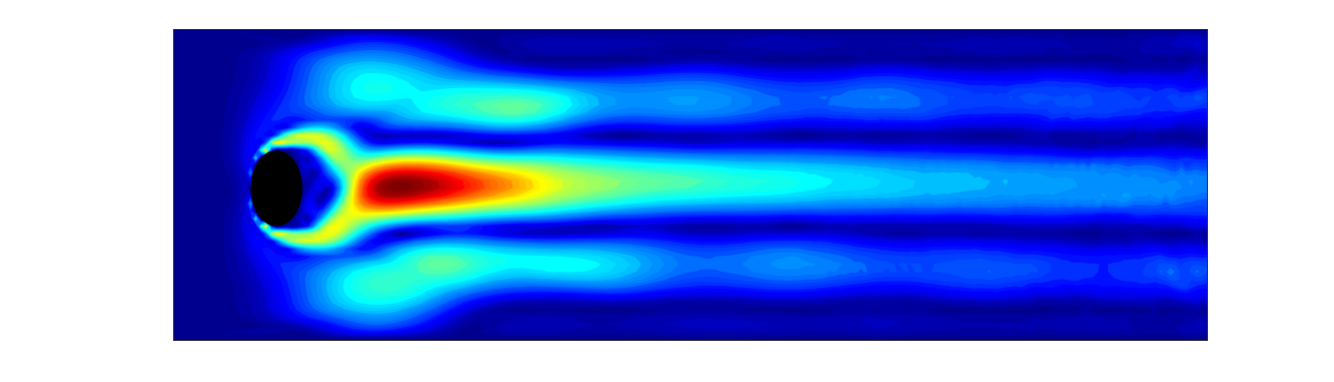}\\
	\includegraphics[width = .32\textwidth, height=.13\textwidth,viewport=80 20 580 170, clip]{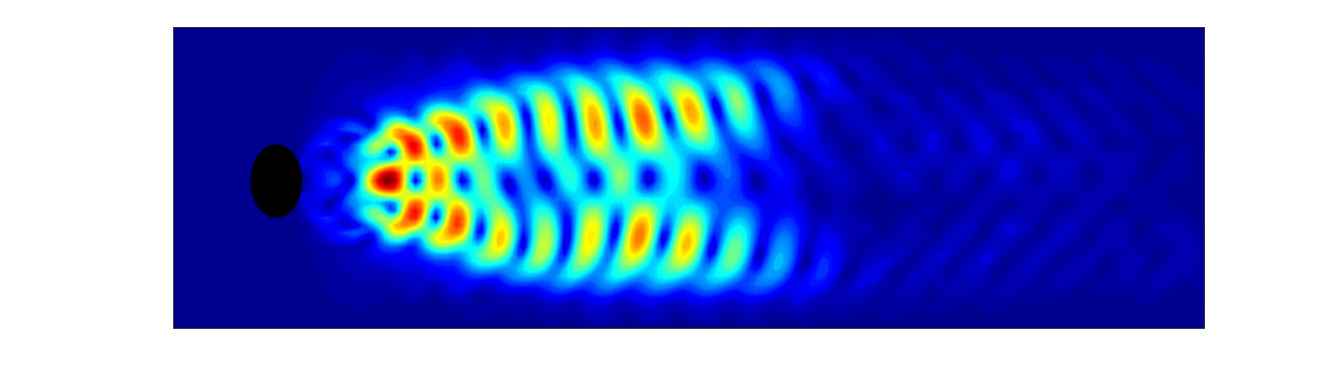}
	\includegraphics[width = .32\textwidth, height=.13\textwidth,viewport=80 20 580 170, clip]{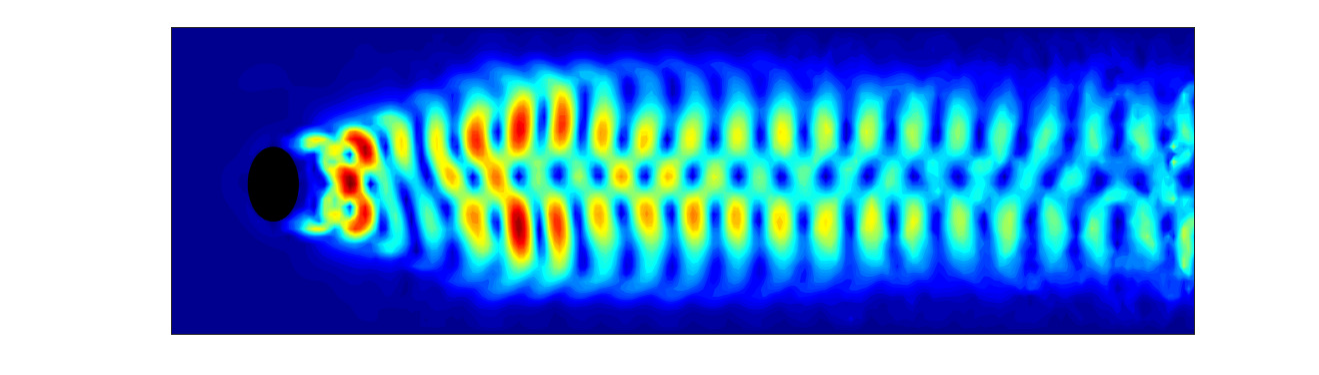}
	\includegraphics[width = .32\textwidth, height=.13\textwidth,viewport=80 20 580 170, clip]{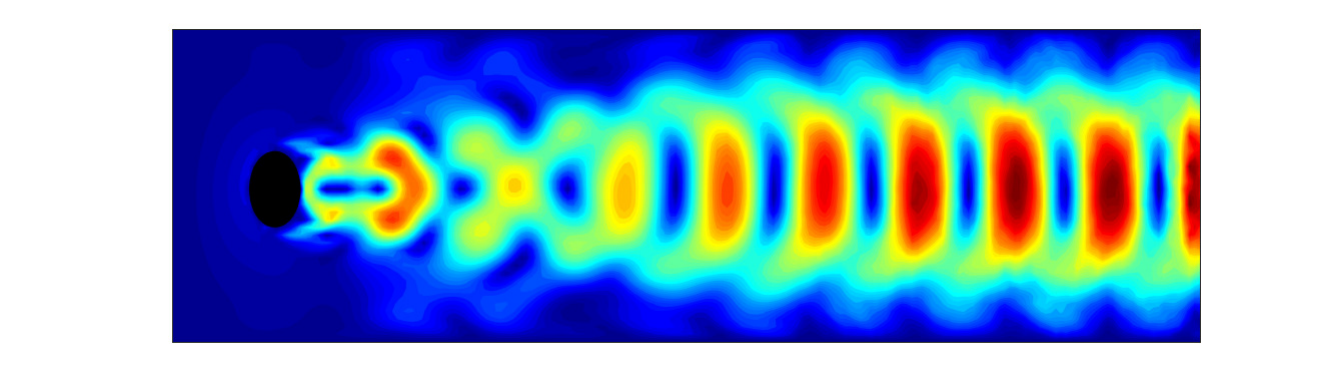}\\
	\includegraphics[width = .32\textwidth, height=.13\textwidth,viewport=80 20 580 170, clip]{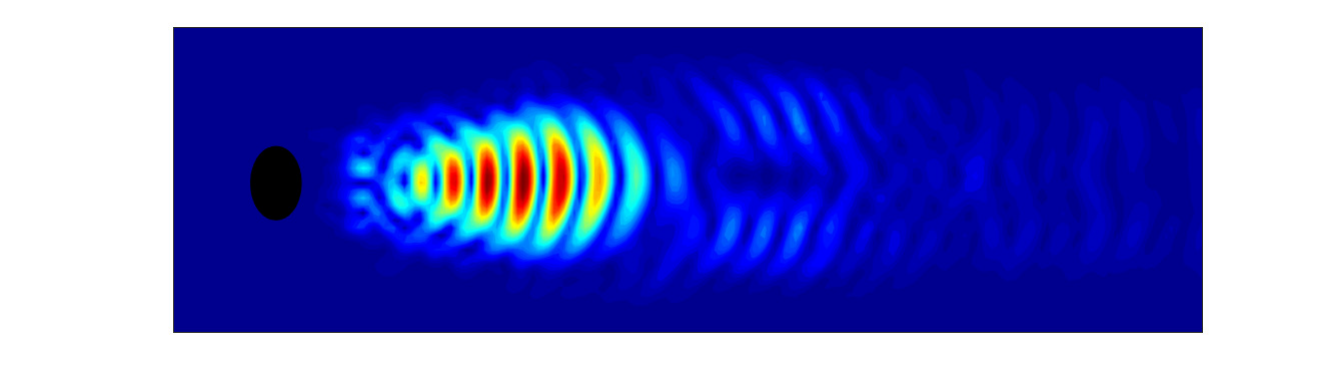}
	\includegraphics[width = .32\textwidth, height=.13\textwidth,viewport=80 20 580 170, clip]{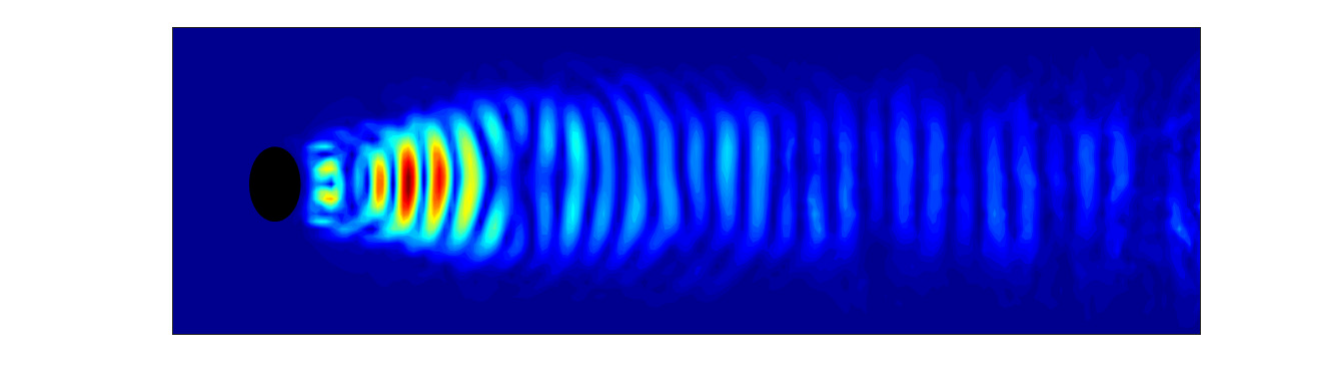}
\	\includegraphics[width = .32\textwidth, height=.13\textwidth,viewport=80 20 580 170, clip]{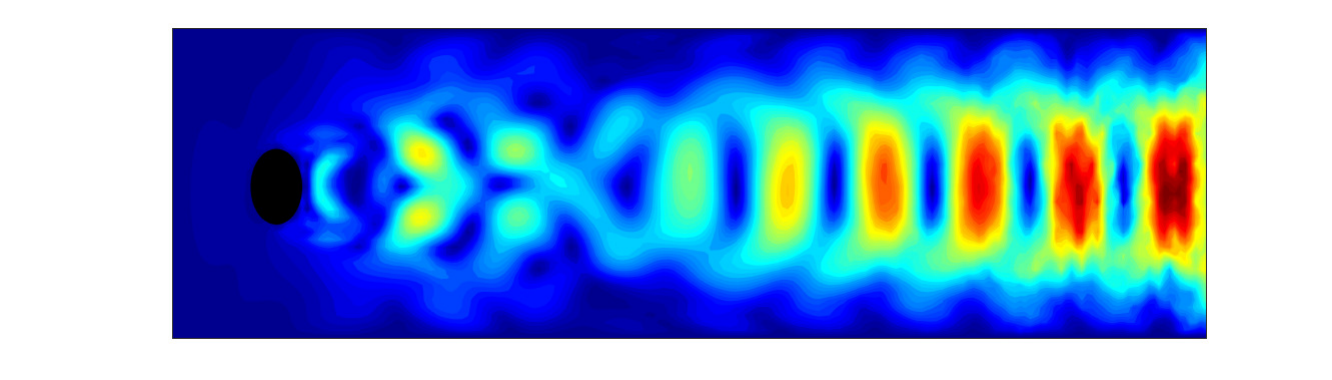}\\

	\caption{\label{modes} Modes 2,3,4,5,6,7 (from top to bottom) for (left) Re=110 POD-ROM, (center) Re=380 POD-ROM, and (right) universal basis}
\end{figure}

}

\subsection{ROM accuracy}

\begin{table}[h]
	\caption{Relative $L^2$ norms of the errors between FOM and ROM solutions for 3 values of $Re$ numbers that are not in the training set. The results are for $\ell=20$. }
	\label{tab:err1}
	\small
	\begin{center}
		\begin{tabular}{c| cc| cc | cc}
			\hline
			& \multicolumn{2}{c|}{Re=30} 	& \multicolumn{2}{c|}{ Re={110}} 	& \multicolumn{2}{c}{ Re=380}\\
			K     & 13  &25 &      13  &25 & 13  &25\\
			\hline\\[-2ex]
	interp. LRTD--ROM	     &4.1e-7 & 2.1e-8 & 1.7e-3 & 1.7e-3& 1.9e-2 & 1.6e-2\\
	non-interp. LRTD--ROM	 &1.6e-6 & 2.2e-7 & 2.8e-3 & 2.2e-3& 1.8e-2& 1.5e-2\\
	POD--ROM              	 &7.6e-5 & 5.5e-5 & 5.6e-2 &5.2e-2 & 1.6e-1& 9.0e-2\\
	\hline
		\end{tabular}
	\end{center}
\end{table}

{ We next study the dependence of the LRTD--ROMs' solution accuracy on the parameter sampling and the ROM design. 
We consider two  training sets with $K=13$ and $K=25$ viscosity parameters log-uniformly sampled in the parameter domain, i.e. $\nu_i=\nu_{\min}\big(\frac{\nu_{\max}}{\nu_{\min}}\big)^{(i-1)/K}$ for {\small $i=1,\dots,K$}. Other parameters of the ROMs were $\eps=10^{-4}$ and $\ell=20$ (dimension of the ROM space).
 We run ROM simulations for $Re=100$, which corresponds to $\nu=10^{-3}$ from the training sets, and for $Re\in\{30,\, 110,\, 380\}$, which are viscosity parameters not in the training sets. 
  For the initial flow velocity we use linear interpolation between known velocity values at the two closest points from the training set at  $t=5$.

Table~\ref{tab:err1} shows the relative $L^2((5,6)\times\Omega )$ error in three different norms for both interpolatory and non-interpolatory versions of the LRTD--ROM (for both $K=13$ and $K=25$) and compares both to the POD--ROM.  We observe that the POD--ROM is worse in all cases, often by an order of magnitude or more.  The interpolatory LRTD--ROM is somewhat more accurate in this measure than the non-interpolatory one for $Re=110$ and $Re=30$ but has similar accuracy at $Re=380$.

}

\begin{figure}[h]\caption{\label{fig:Re100} Prediction of lift and  drag coefficients  for Re=100 (which is in the training set). Number of parameters in the training set is $K=13$, and $\ell=20$.}
	\centering
	\includegraphics[width=0.4\textwidth]{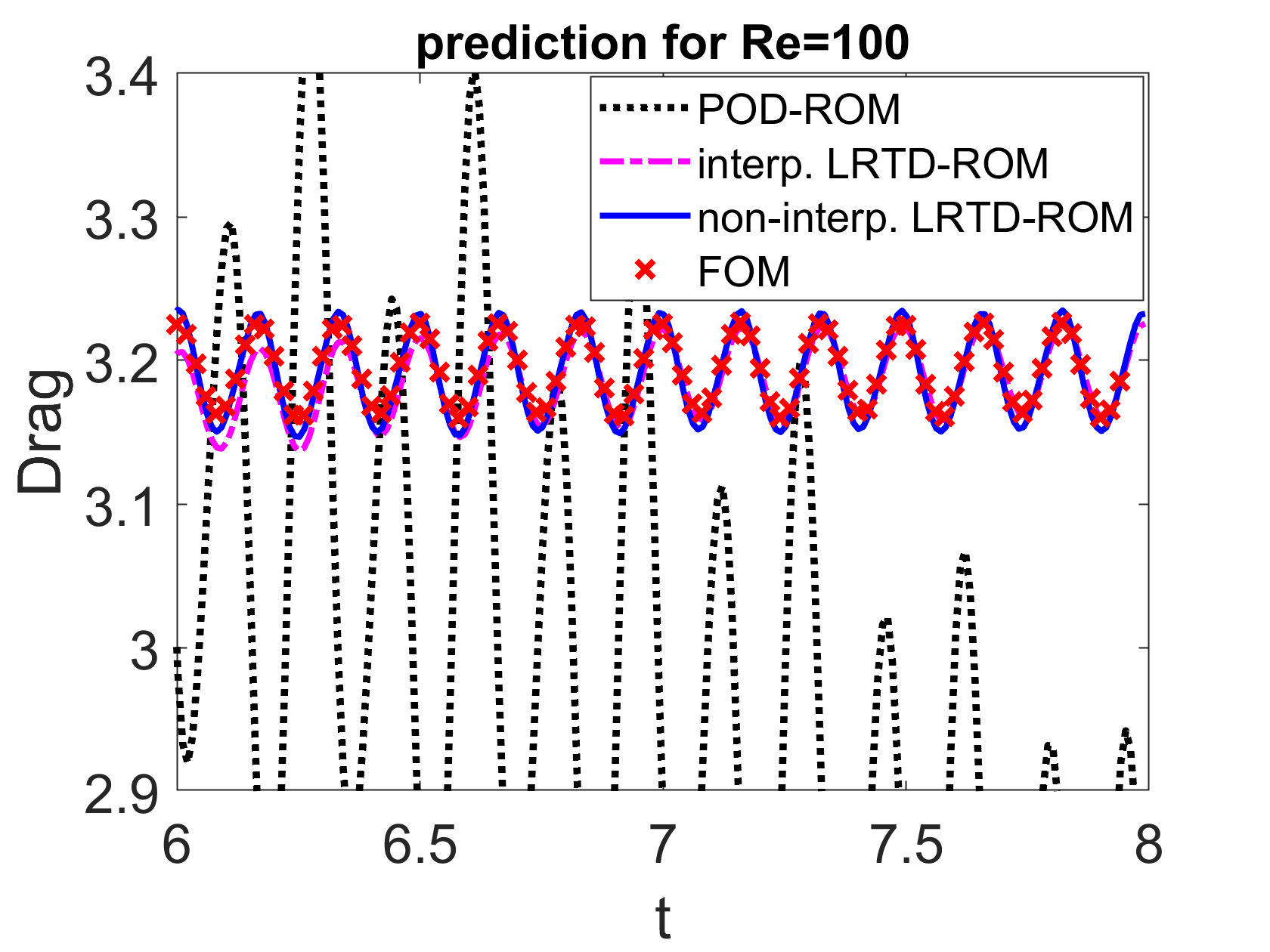}  \qquad\includegraphics[width=0.4\textwidth]{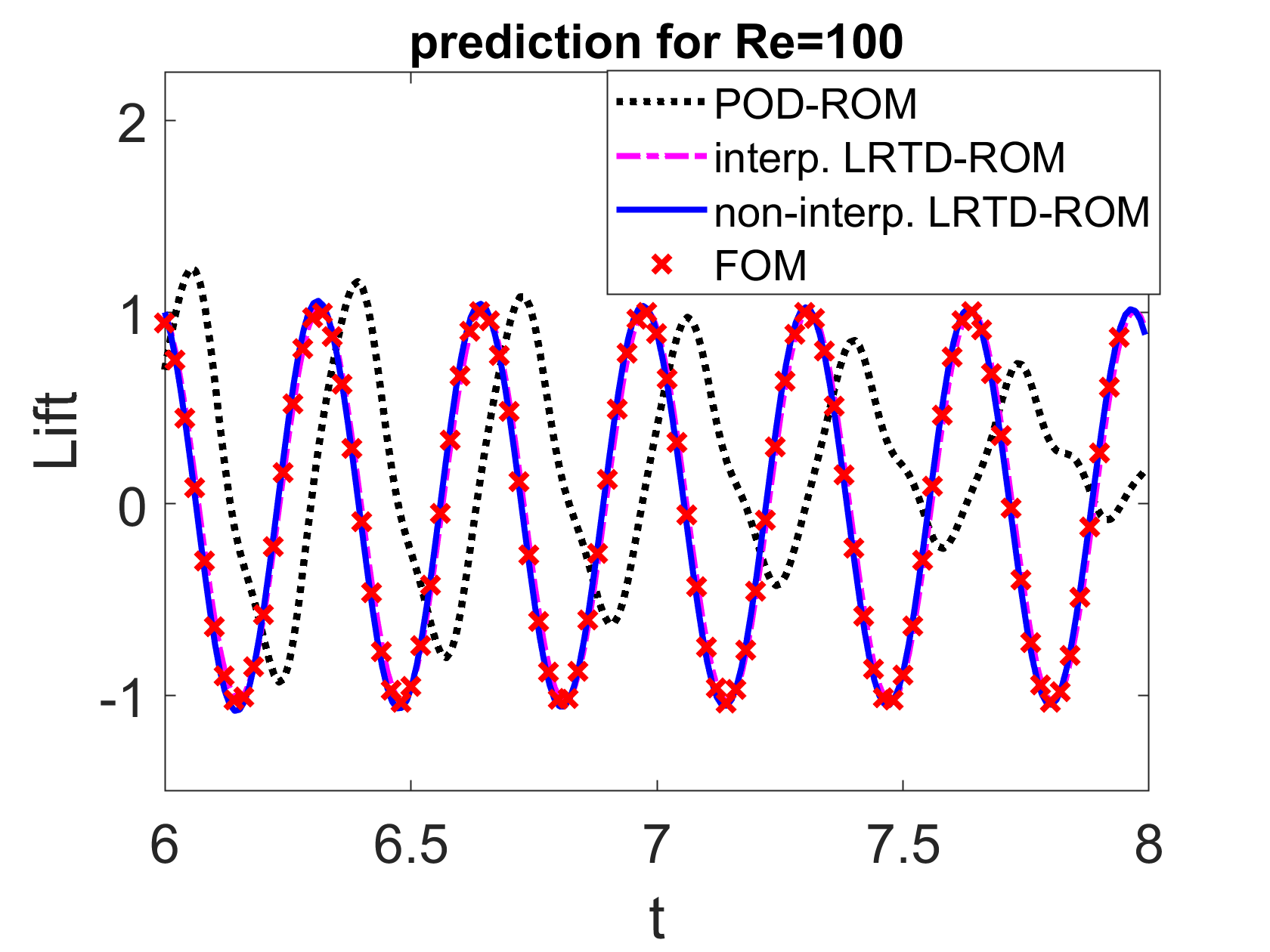}
\end{figure}

\begin{figure}[h]\caption{\label{fig:ReOut1} Prediction of lift and  drag coefficients  for Re=110 and  Re=380 \textbf{not} from the training set. Number of parameters in the training set is $K=13$, and $\ell=20$.}
	\centering
	\includegraphics[width=0.4\textwidth]{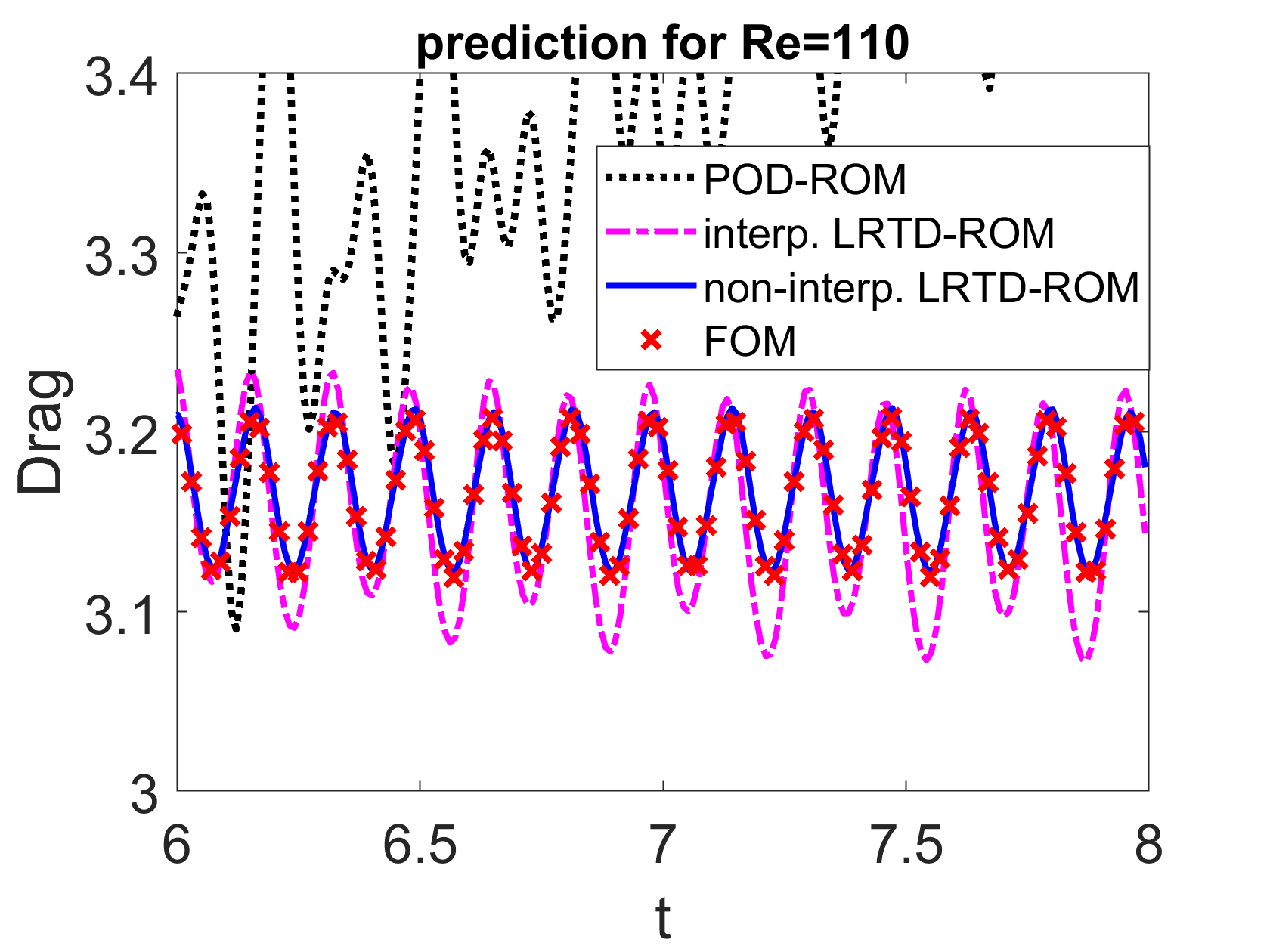}  \qquad\includegraphics[width=0.4\textwidth]{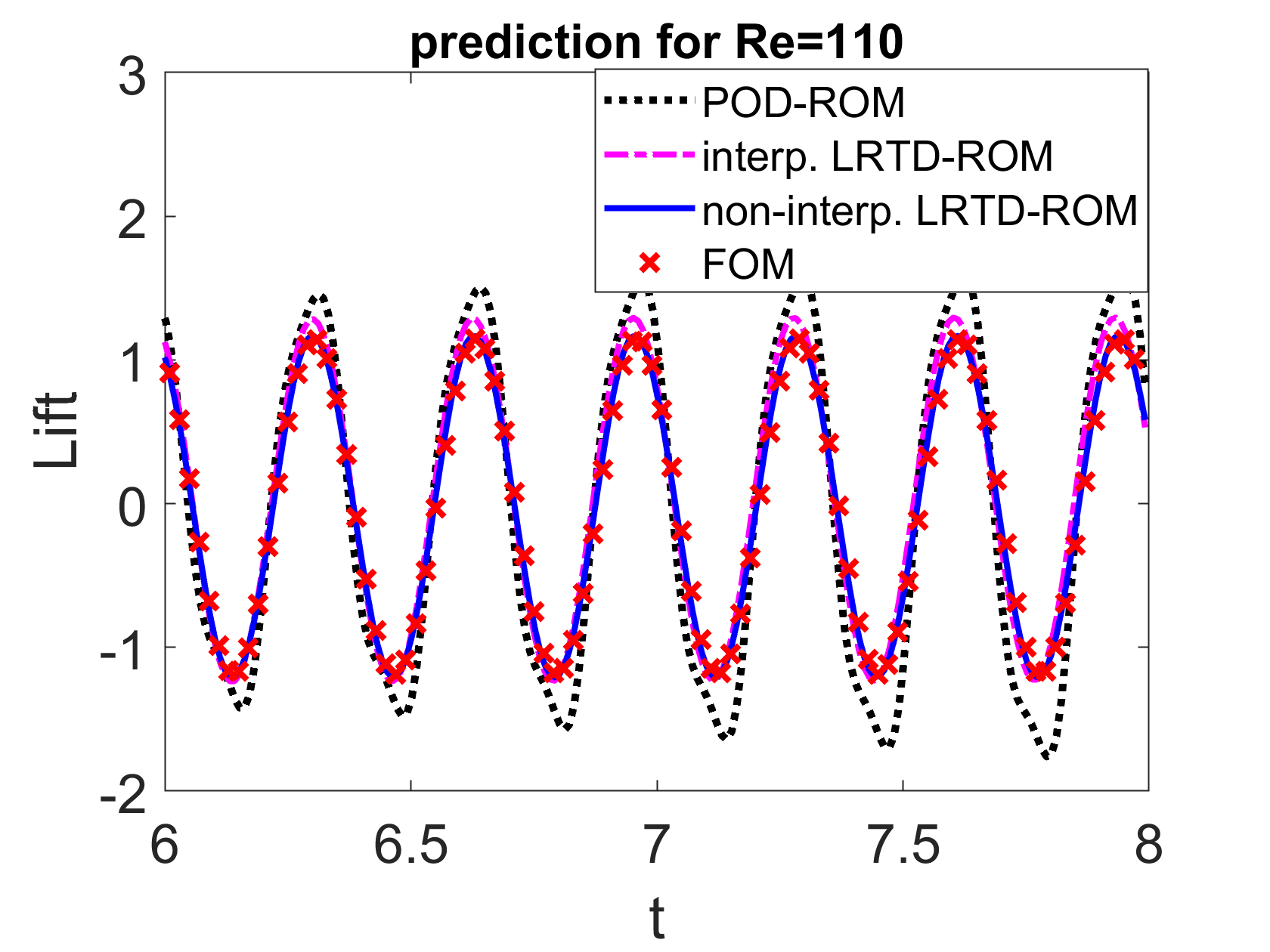}\\
		\includegraphics[width=0.4\textwidth]{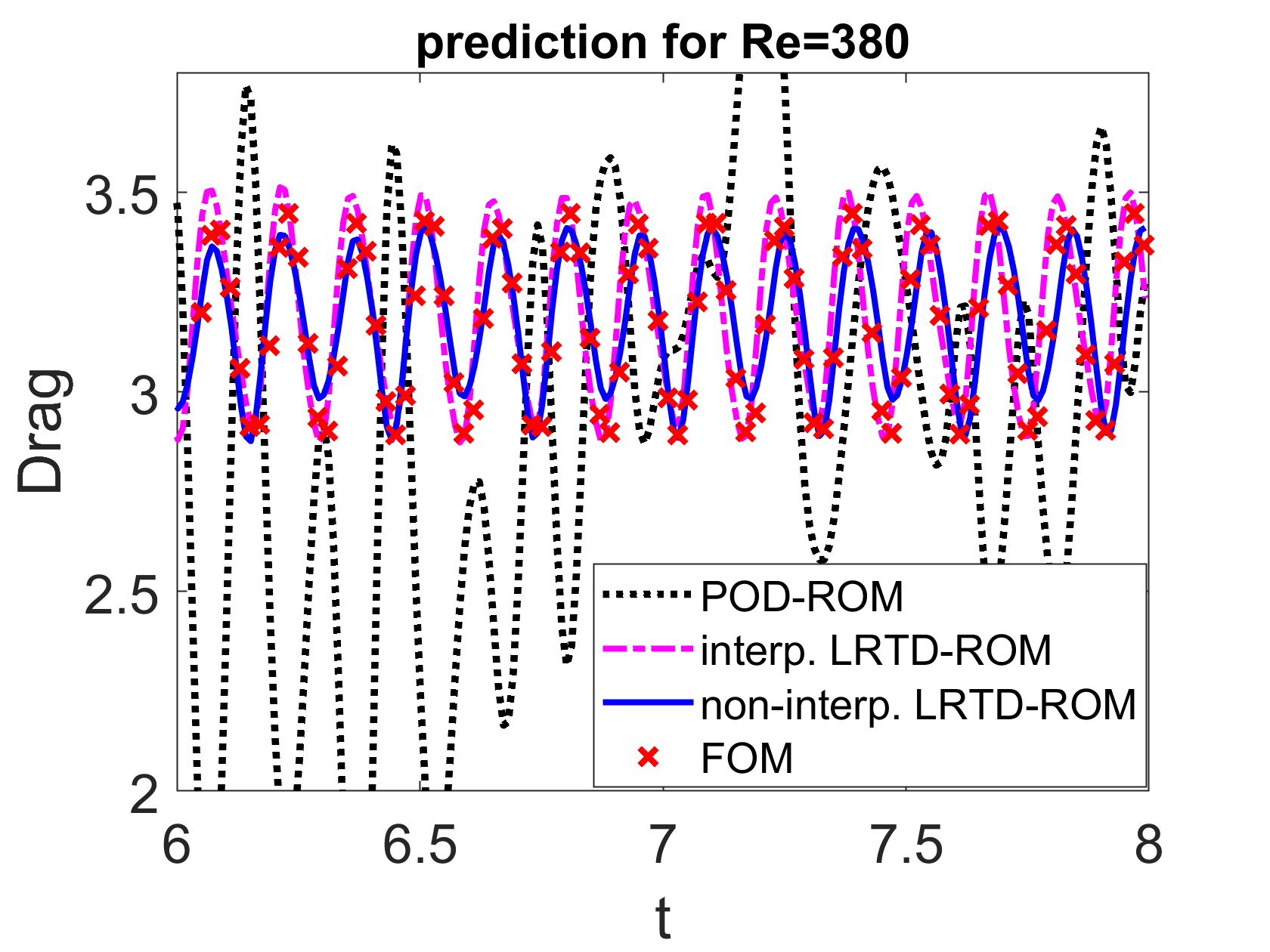}  \qquad\includegraphics[width=0.4\textwidth]{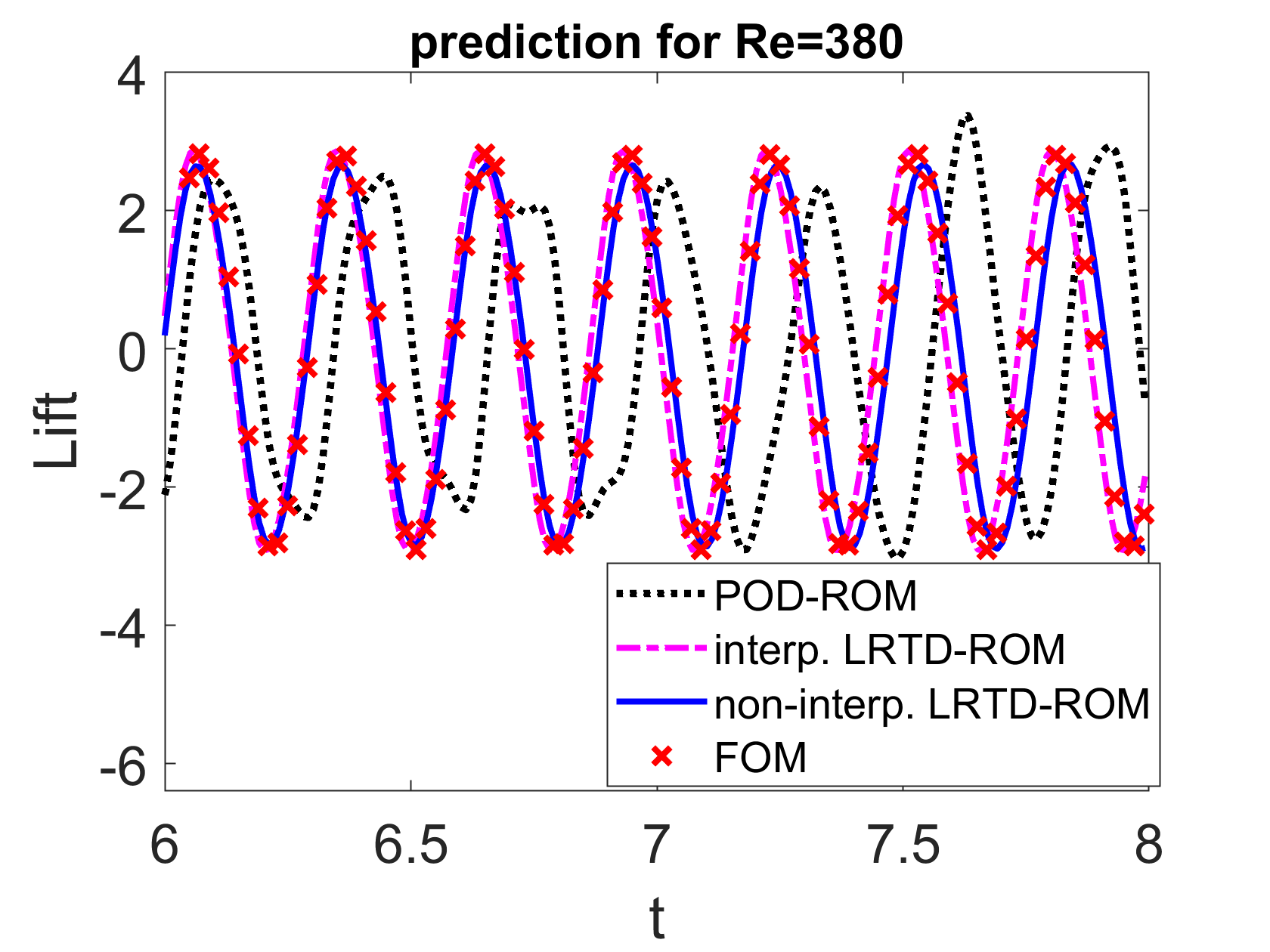}
\end{figure}

\begin{figure}[h]\caption{\label{fig:ReOut2} Prediction of lift and  drag coefficients  for Re=110 and  Re=380 \textbf{not} from the training set. Number of parameters in the training set is $K=25$, and $\ell=20$.}
	\centering
	\includegraphics[width=0.4\textwidth]{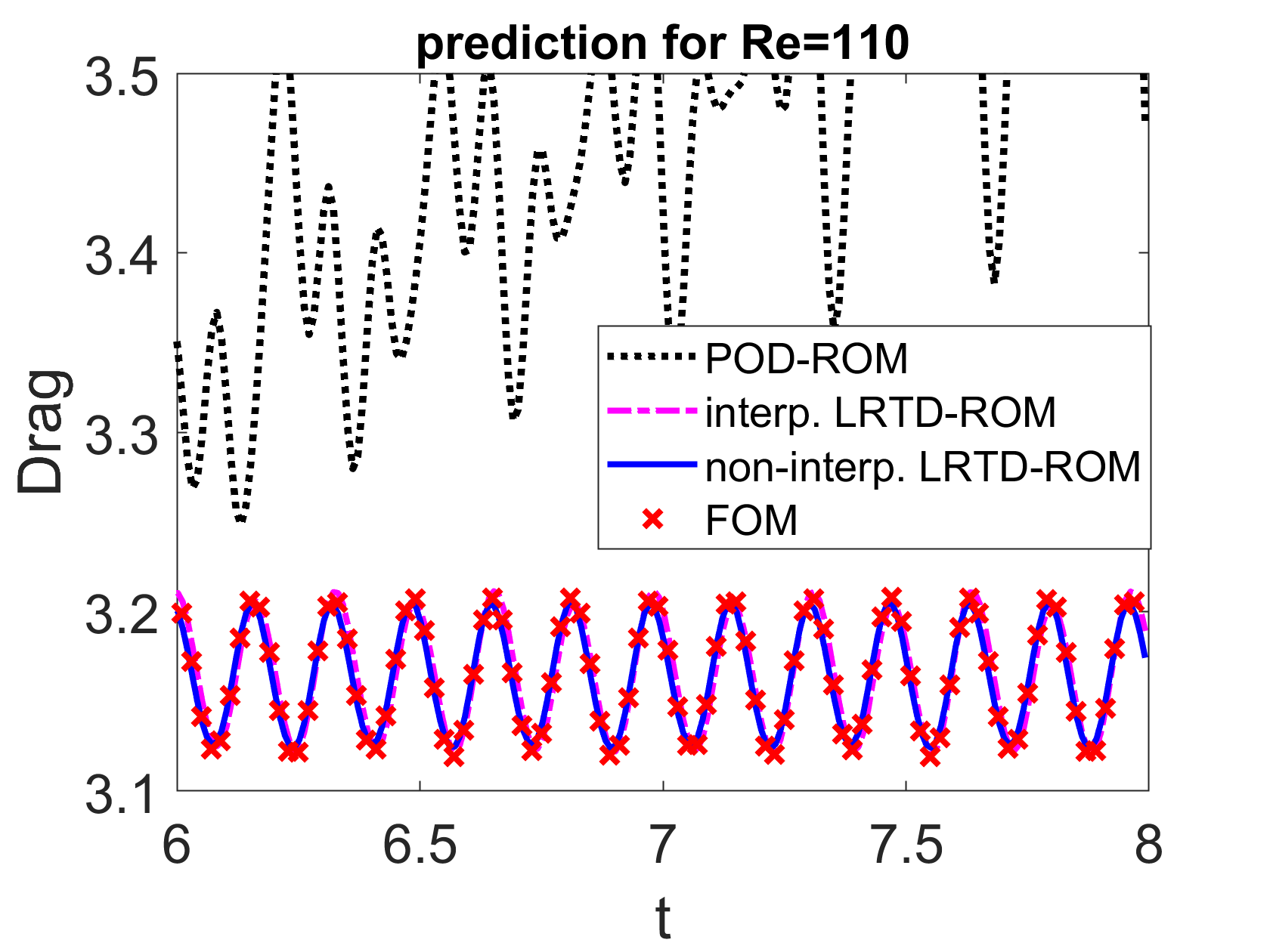}  \qquad\includegraphics[width=0.4\textwidth]{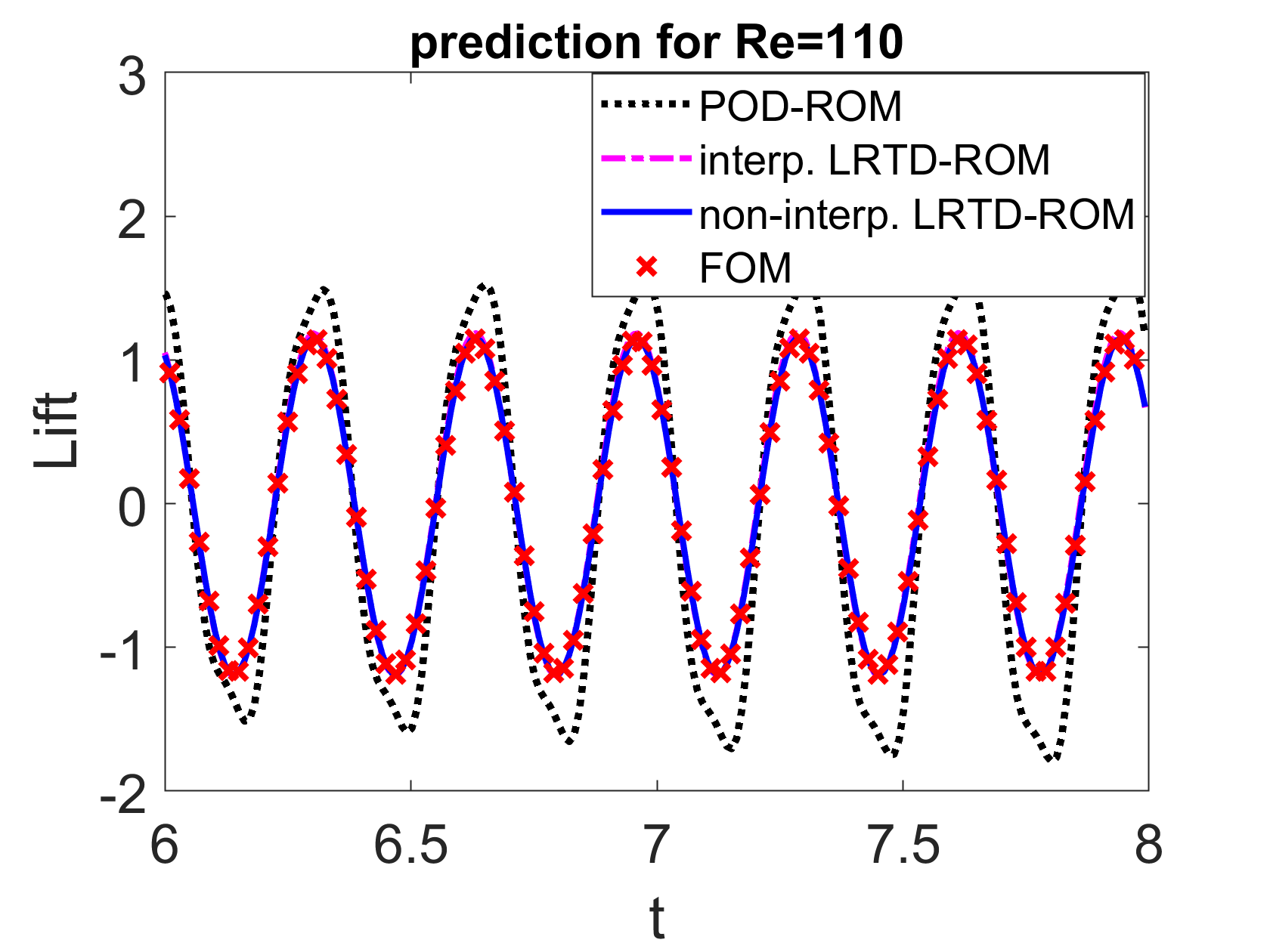}\\
	\includegraphics[width=0.4\textwidth]{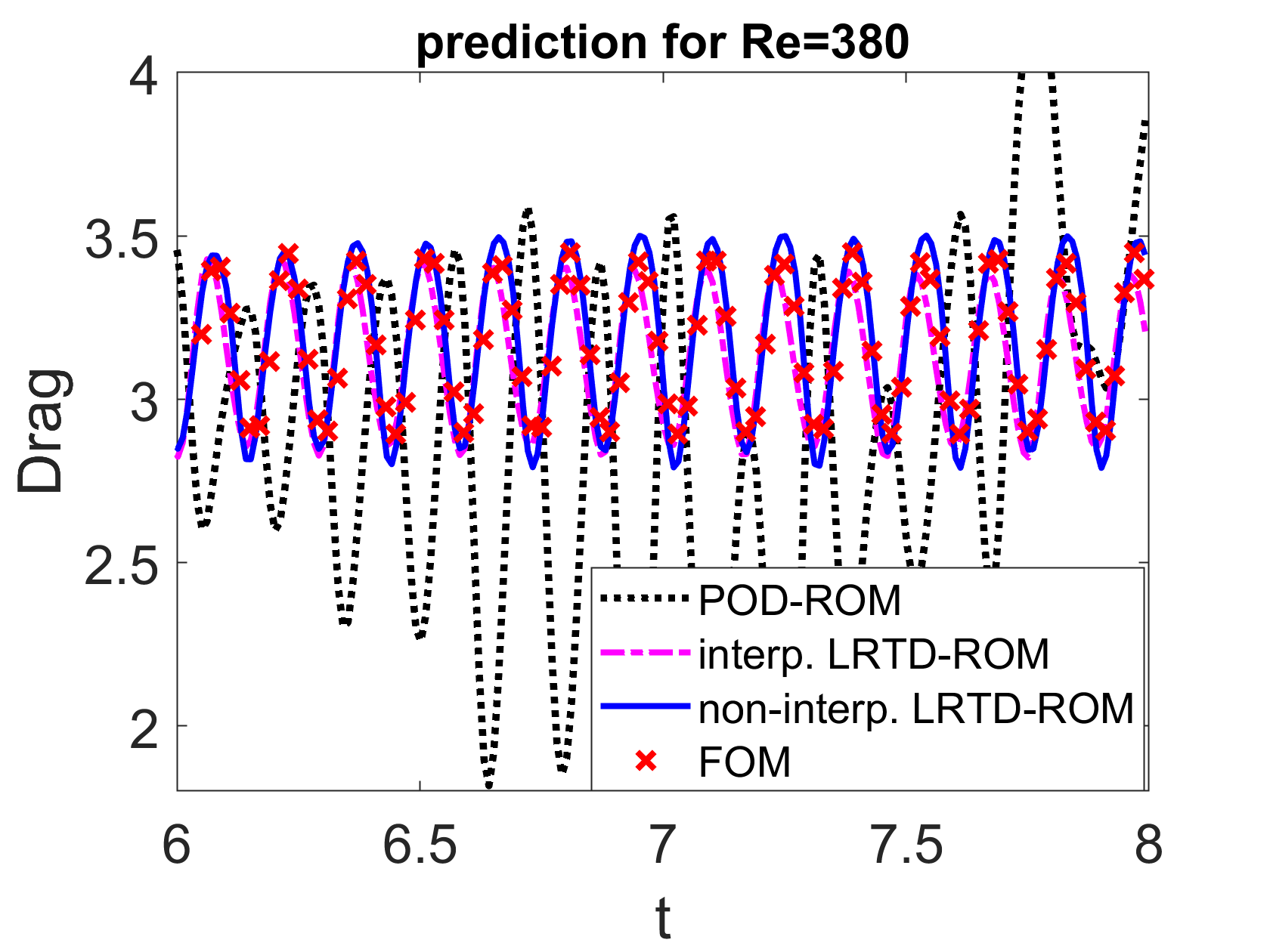}  \qquad\includegraphics[width=0.4\textwidth]{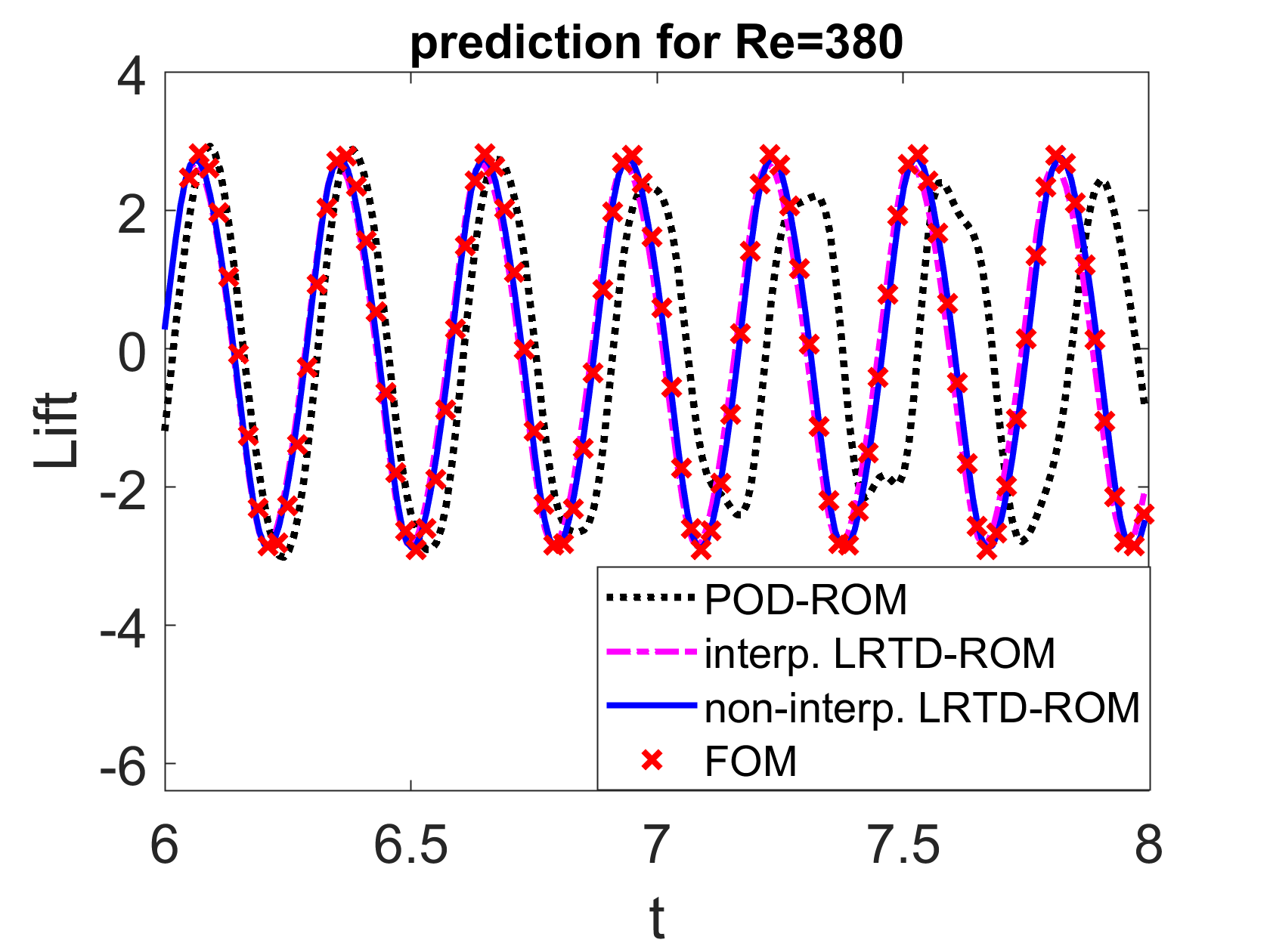}
\end{figure}

In addition to accuracy in norms, we are interested in the ability of the tensor ROMs to predict critical statistics such as drag and lift coefficients for the cylinder.  We are interested in the prediction accuracy of the method both outside the training set and beyond the time interval used to collect the snapshots. 
  The results for  $Re=100$ and $K=13$ are shown in Figure~\ref{fig:Re100}.  We note  that $Re=100$ is in the training set, and observe that the interpolatory and non-interpolatory LRTD-ROM results  were quite good, and match the FOM flow lift and drag quite well.  The POD-ROM results, however, were very inaccurate.  As discussed above, the POD-ROM may need many more modes to be able to capture finer scale detail that the LRTD-ROMs are able to capture.

The results for $Re=110$  and $Re=380$ are presented in Figure~\ref{fig:ReOut1} for $13$  parameters in the training set and in Figure~\ref{fig:ReOut2} for $25$  parameters in the training set.  The plots are for $6\le t \le 8$, since we are not starting with the ``correct'' flow state and the system may take some time to reach the quasi-equilibrium (periodic) state.  We observe that for both $K=13$ and $K=25$, POD-ROM results are inaccurate.  For $K=25$, both interpolatory and non-interpolatory results are reasonably accurate, although for $Re=380$ the drag predictions show some slight error.  For $K=13$, results are less accurate; in the latter case, we observe non-interpolatory  LRTD-ROM results to be slightly better compared to interpolatory LRTD-ROM (similar accuracy is found for total kinetic energy plots, which are omitted).
 
%
%
%
 
  \begin{figure}[h]\caption{\label{fig:ell} Prediction of drag and lift coefficients for $Re=380$ with non-interpolatory LRTD--ROM dimensions $\ell=20$ and $\ell=30$. Number of parameters in the training set is $K=25$.}
 	\centering
 	\includegraphics[width=0.4\textwidth]{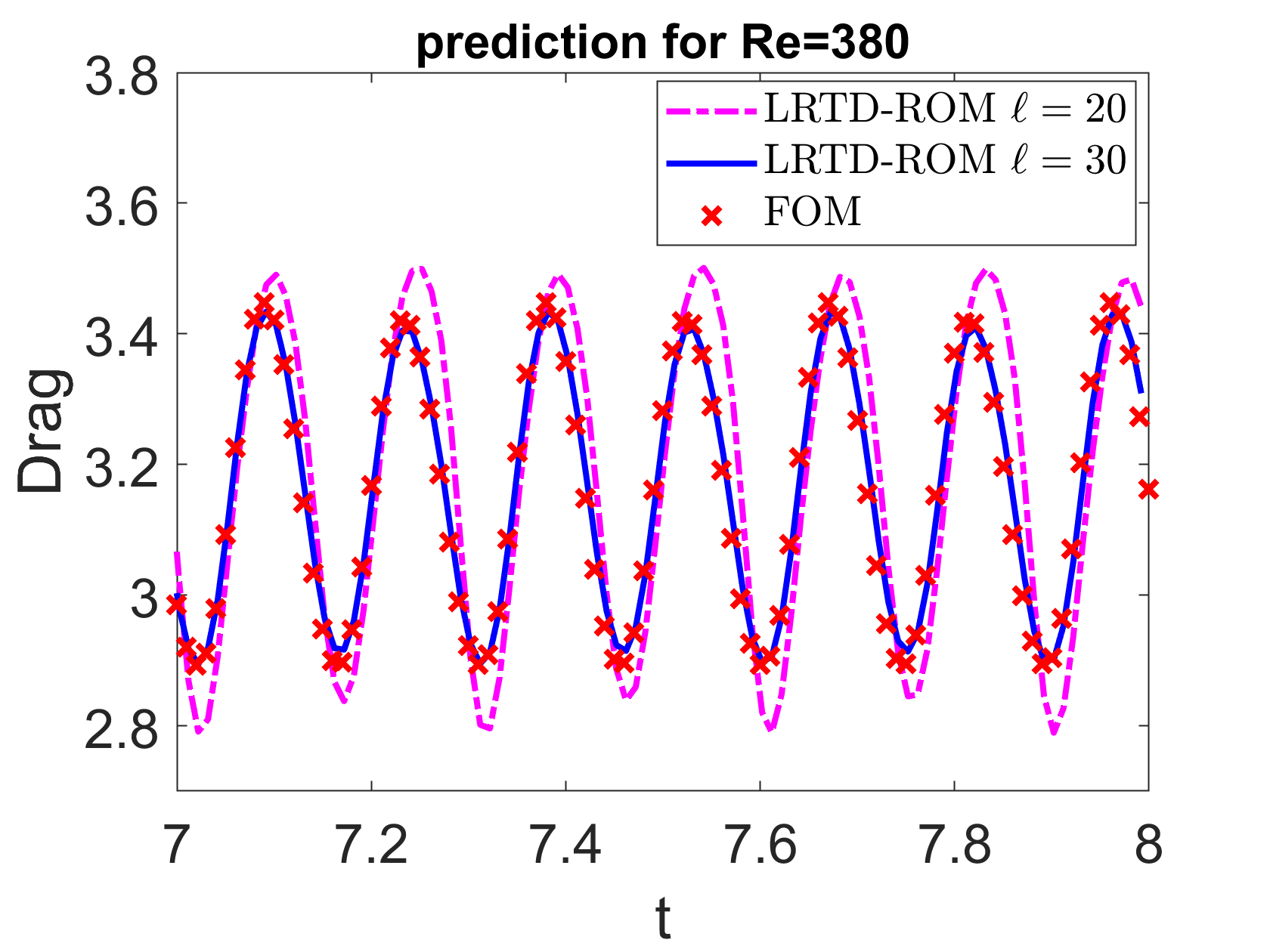}\qquad\includegraphics[width=0.4\textwidth]{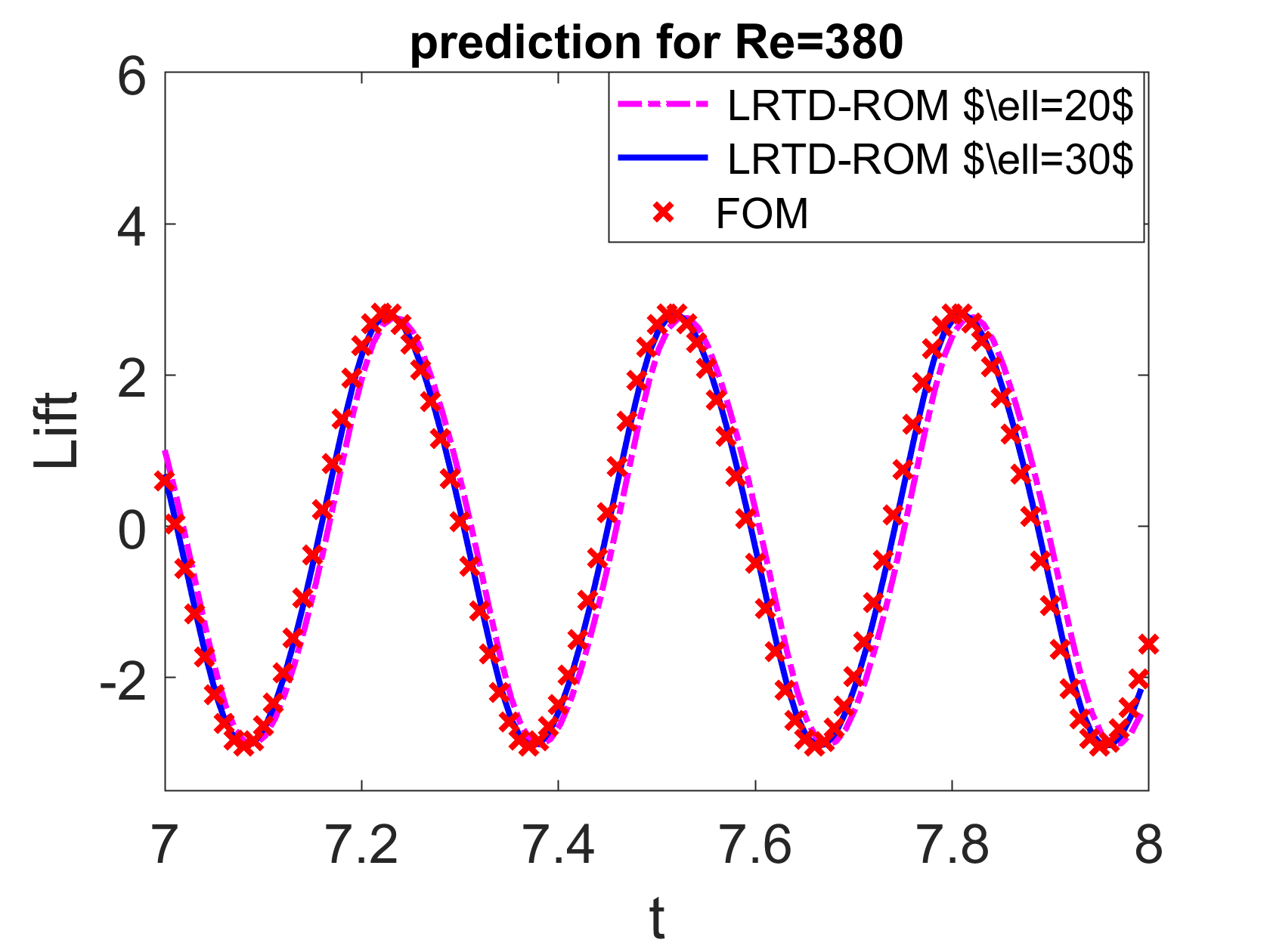}
 \end{figure}
 
Increasing the dimension of the LRTD--ROM improves the accuracy, as should be expected. The effect is illustrated in figure~\ref{fig:ell}, which shows the results for the non-interpolatory LRTD--ROM with $\ell=20$ and $\ell=30$. While the results for the lift prediction are almost indistinguishable, for the drag coefficient $\ell=30$ has the ROM and FOM values match, while $\ell=20$ is seen to slightly overshoot minimal and maximal values for $Re=380$.
Thus for  further simulations we chose the non-interpolatory LRTD--ROM with $\ell=30$, and trained on the set of $25$ parameters.


\begin{figure}[h]\caption{\label{fig:minmax} Minimal and the maximal values of drag and lift coefficients along the smooth branch of solutions. }
	\centering
	\includegraphics[width=0.4\textwidth]{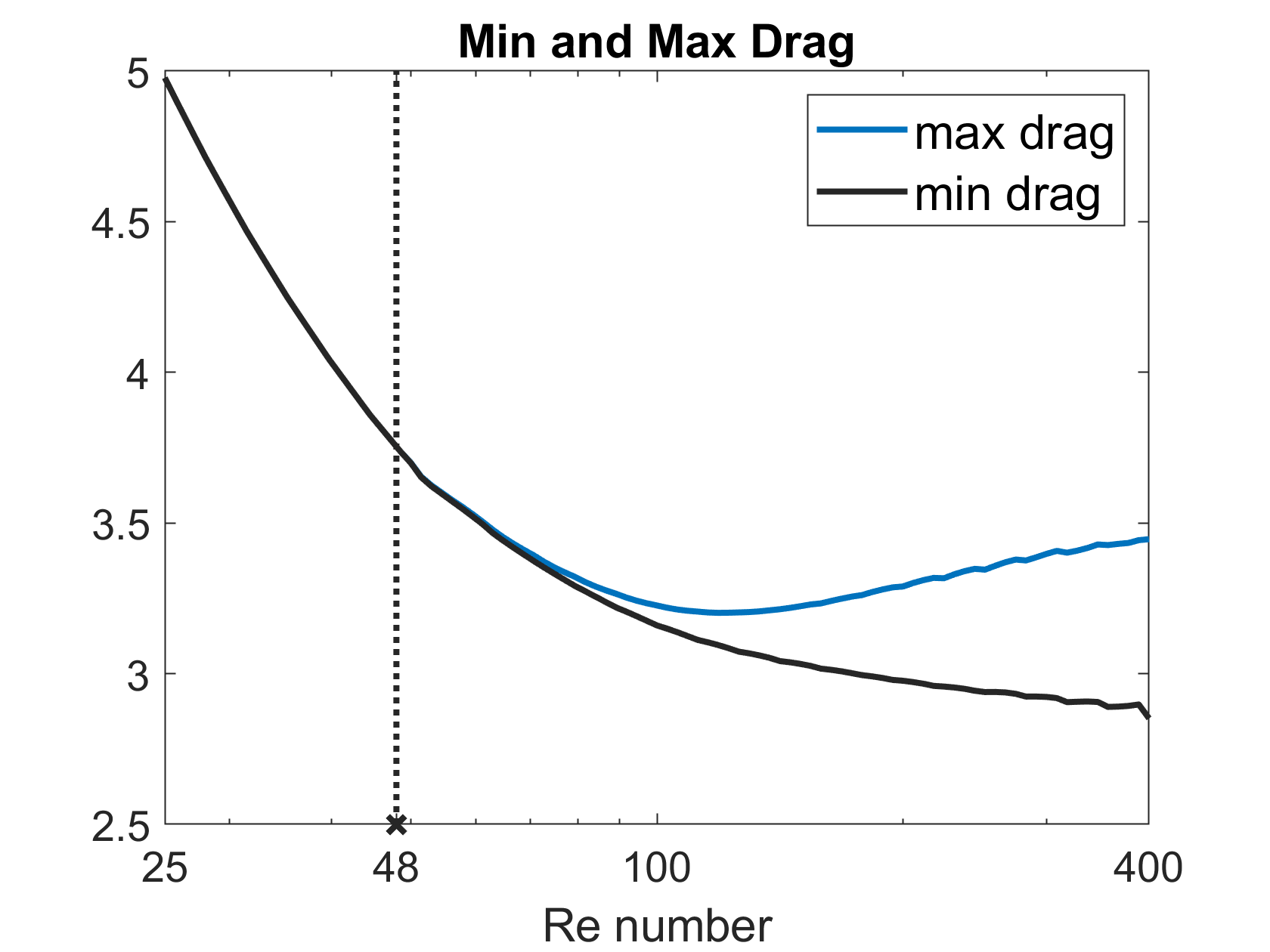}\qquad\includegraphics[width=0.4\textwidth]{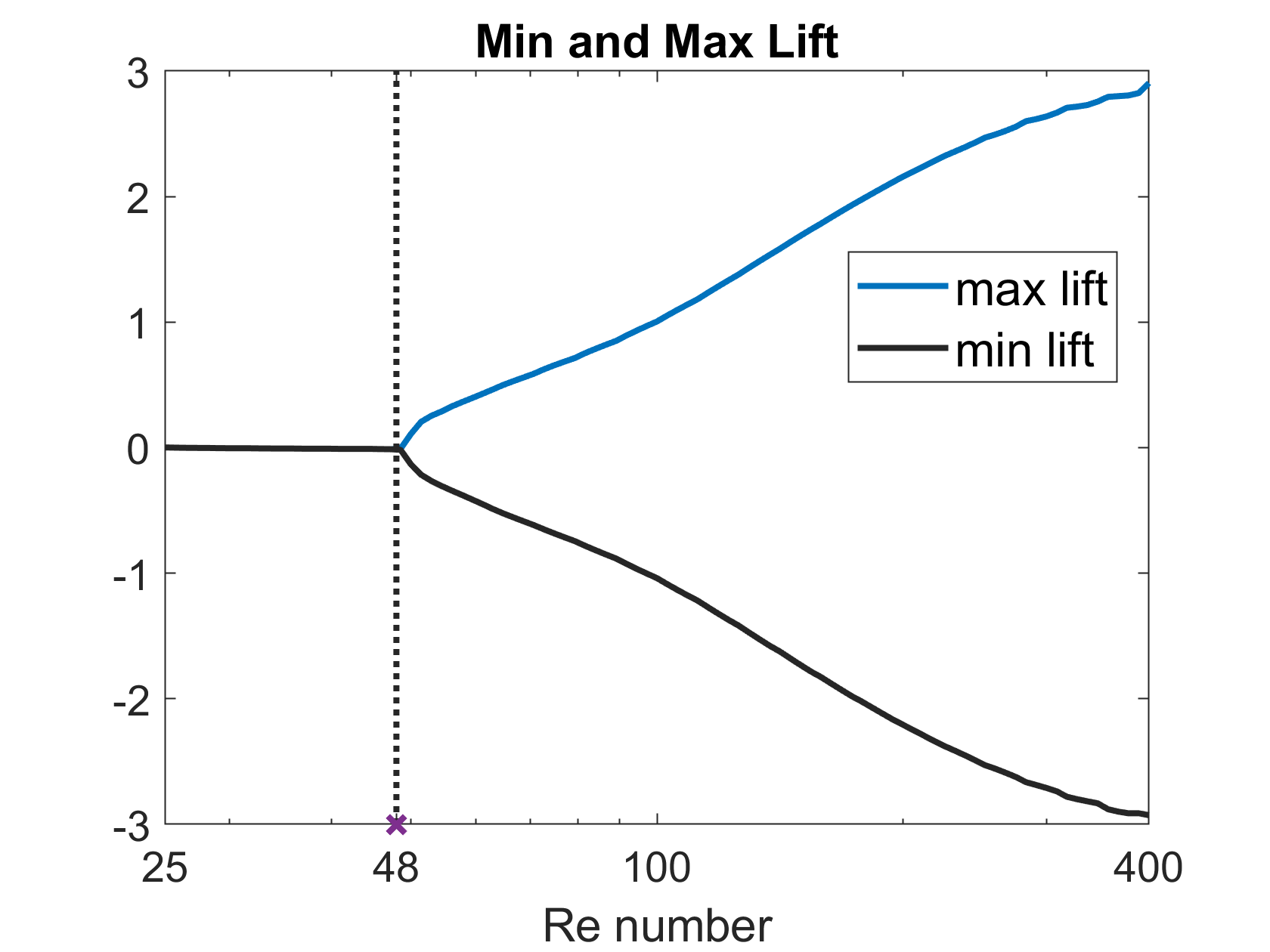}
\end{figure}
 
 \subsection{Predicting an entire branch of solutions}
 
 We are interested in applying the ROM to approximate the flow statistics along the whole branch of solutions, and for these tests we use non-interpolatory LRTD--ROM with $\ell=30$ and $K=25$.   To this end, we run the LRTD-ROM for 99 viscosity values log-uniformly sampled in our parameter domain and calculate the solution up to final $T=50$ starting from an initial condition which is interpolated from snapshots at $t_0=5$. 
 %
 Figure~\ref{fig:minmax} shows the predicted lift and drag coefficient's variation for varying $Re$, after quasi-periodic state is reached in each flow.  We find the transition point from steady-state to periodic flow to be near $Re=48$, which agrees closely with the literature \cite{chen1995bifurcation,williamson1996vortex}.
 
 \begin{figure}[h]\caption{\label{fig:spectrum} Shown below are spectrums of the lift coefficients for varying $Re$.}
 	\centering
 	\includegraphics[width=0.4\textwidth]{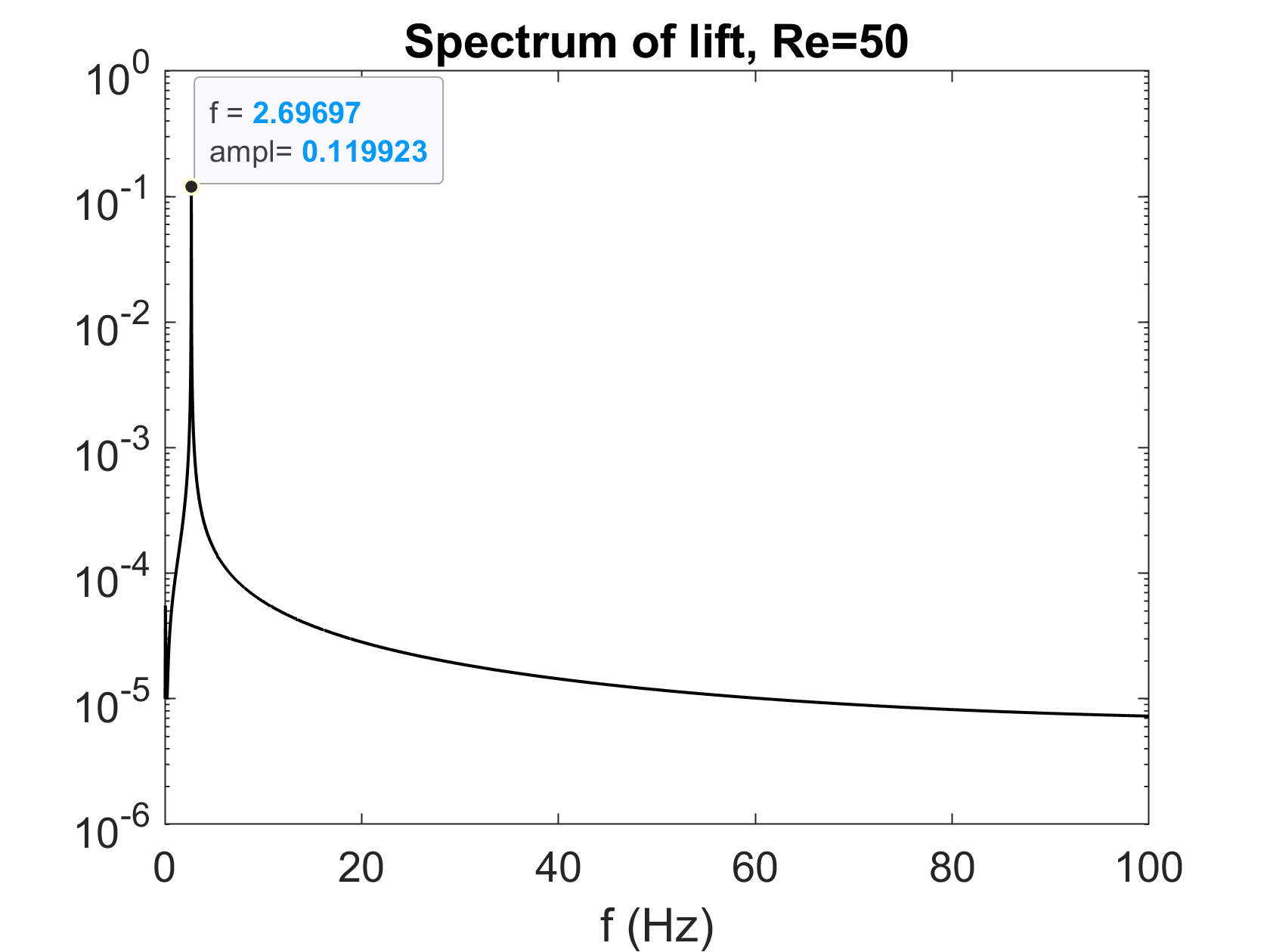}\qquad\includegraphics[width=0.4\textwidth]{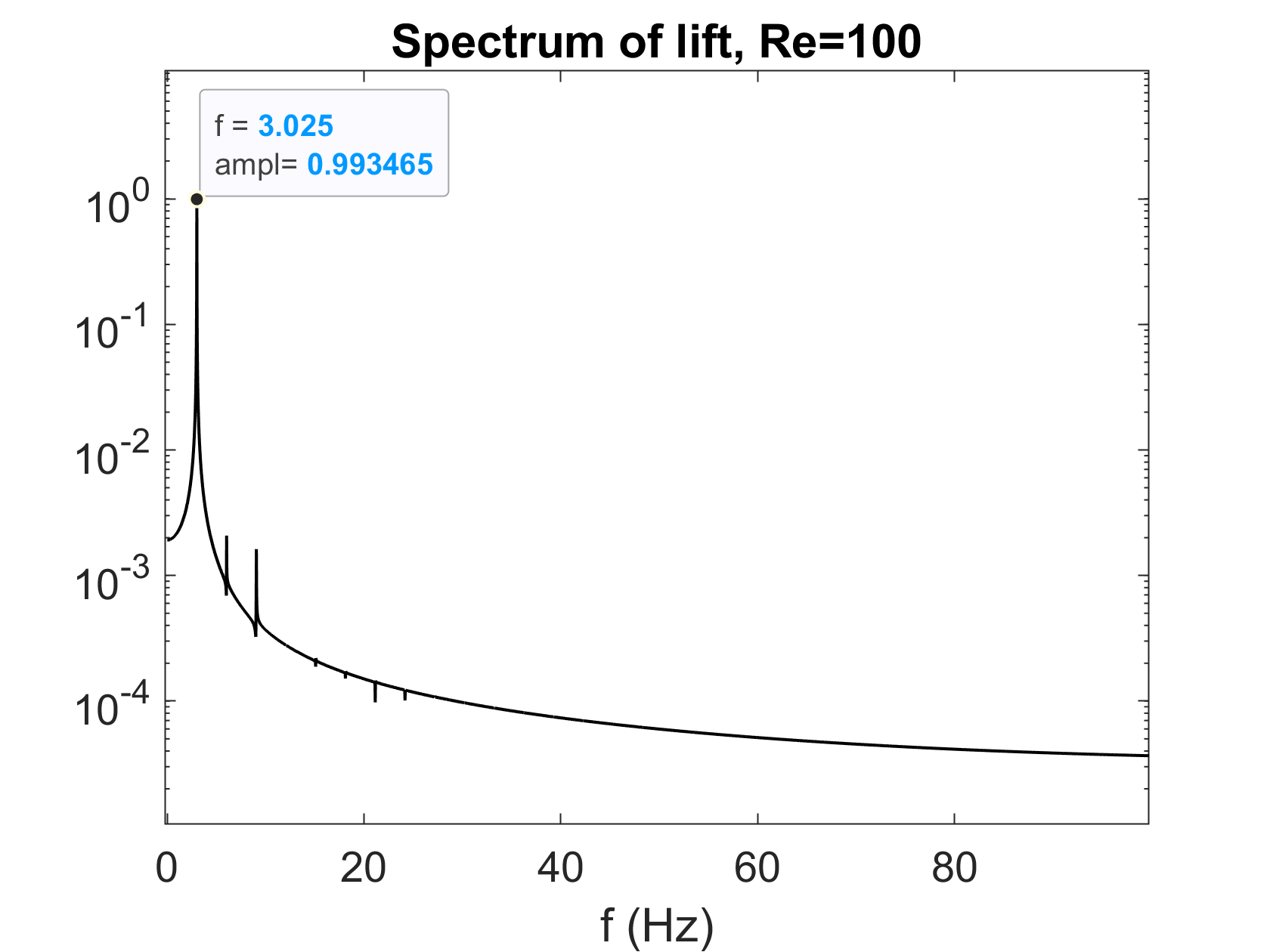}\\
 	 	\includegraphics[width=0.4\textwidth]{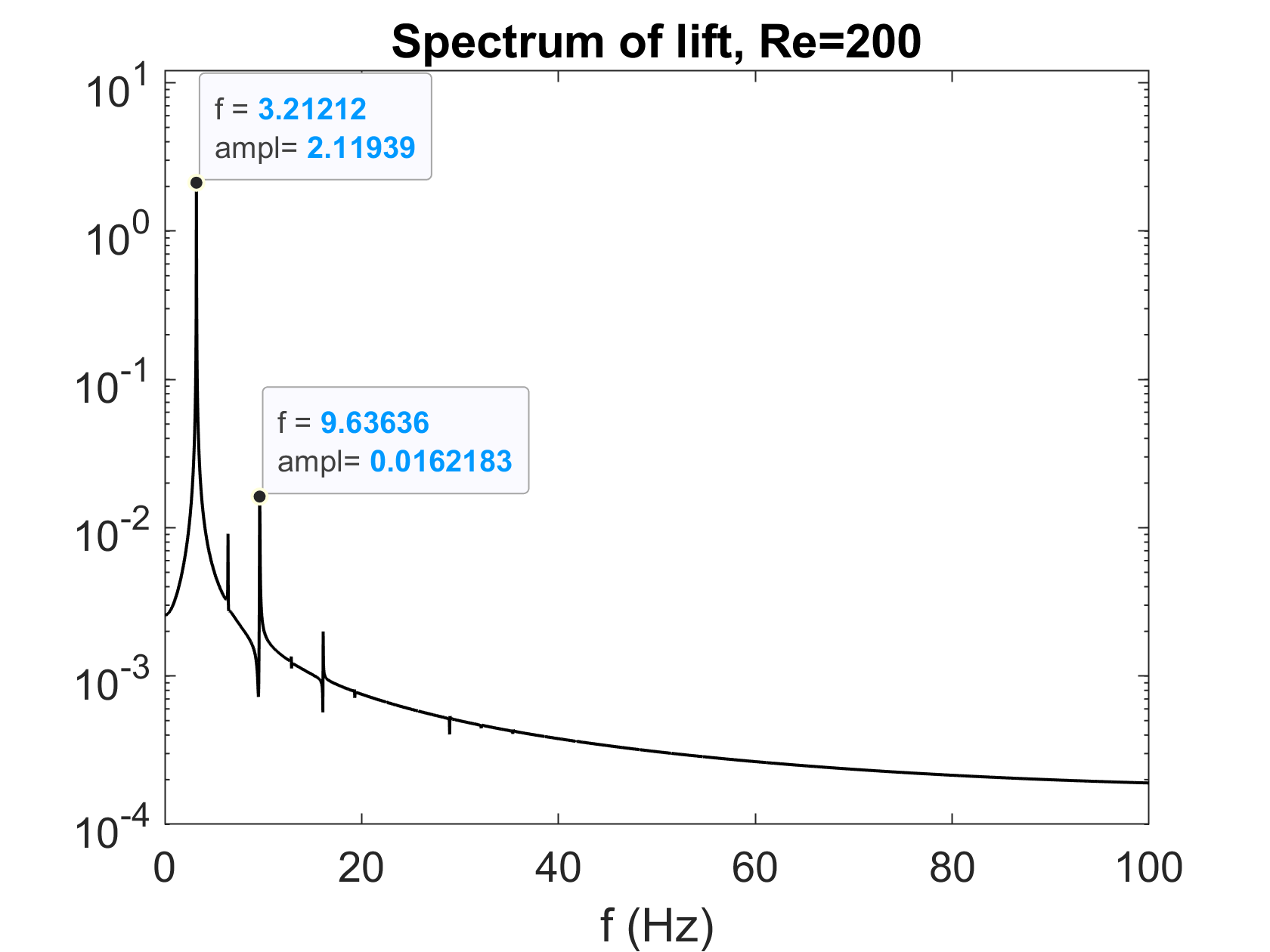}\qquad\includegraphics[width=0.4\textwidth]{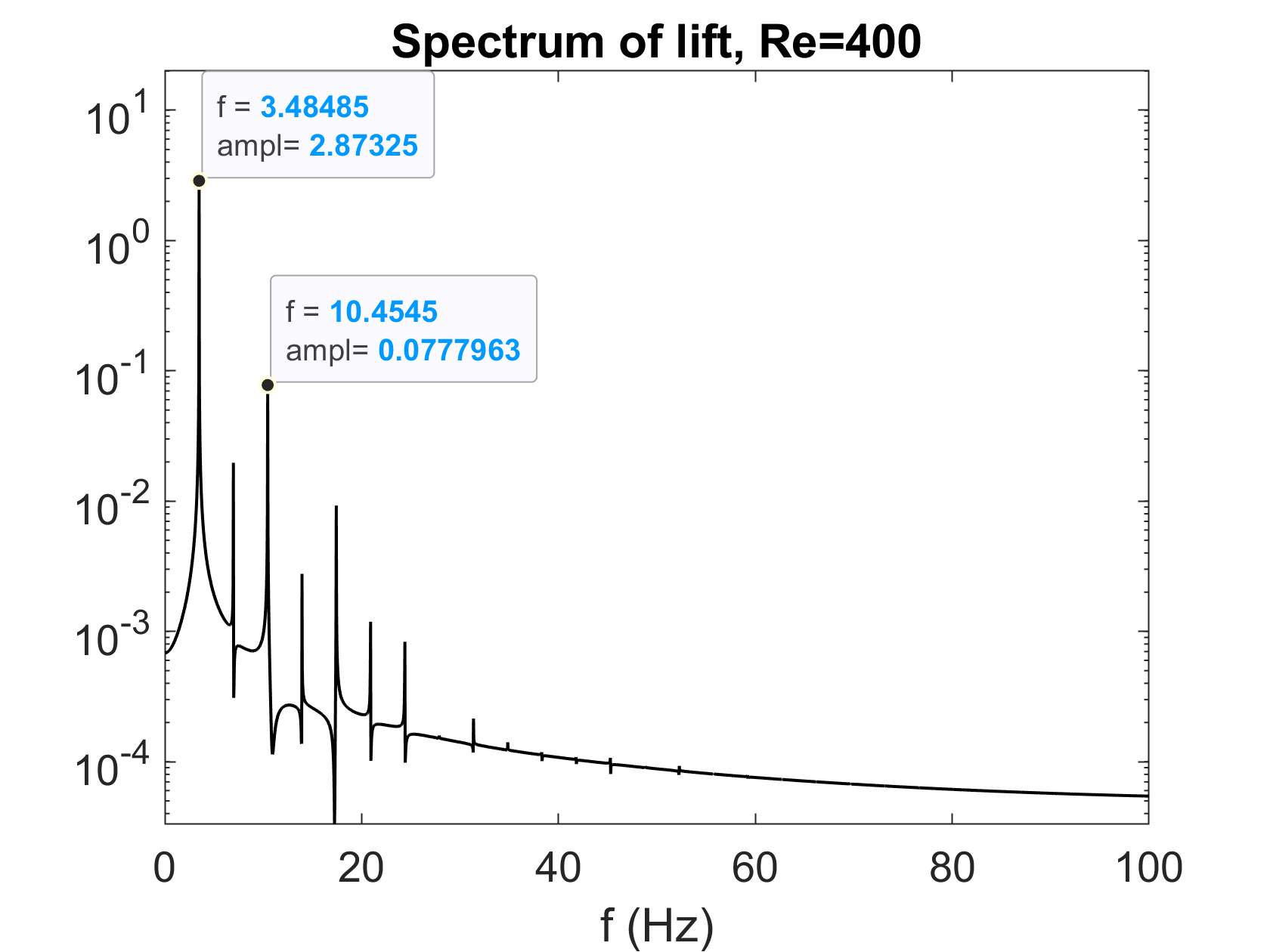}
 \end{figure}
 
 We next consider the spectrum of the flow statistics by computing the Fourier transform of the lift coefficient for time interval $t\in[10,50]$.  In Figure \ref{fig:spectrum}, this is shown for $Re$=50, 100, 200 and 400.  For $Re=50$, only one spike is observed, indicating a single dominant frequency.  For $Re=100$, some smaller spikes are shown in the plot, but they are nearly 3 orders of magnitude smaller than the largest spike and have minimal effect on the solution's periodicity.  By $Re=200$, the second biggest spike is a little over two orders of magnitude smaller than the biggest one, and by $Re=400$ there is less than two orders of magnitude difference suggesting that this flow is moving away from a purely periodic flow to one with more complex behavior in time.
 
%
%

\begin{figure}[h]\caption{\label{fig:modes} First four HOSVD singular vectors in time and parameter modes.}
	\centering
	\includegraphics[width=0.4\textwidth]{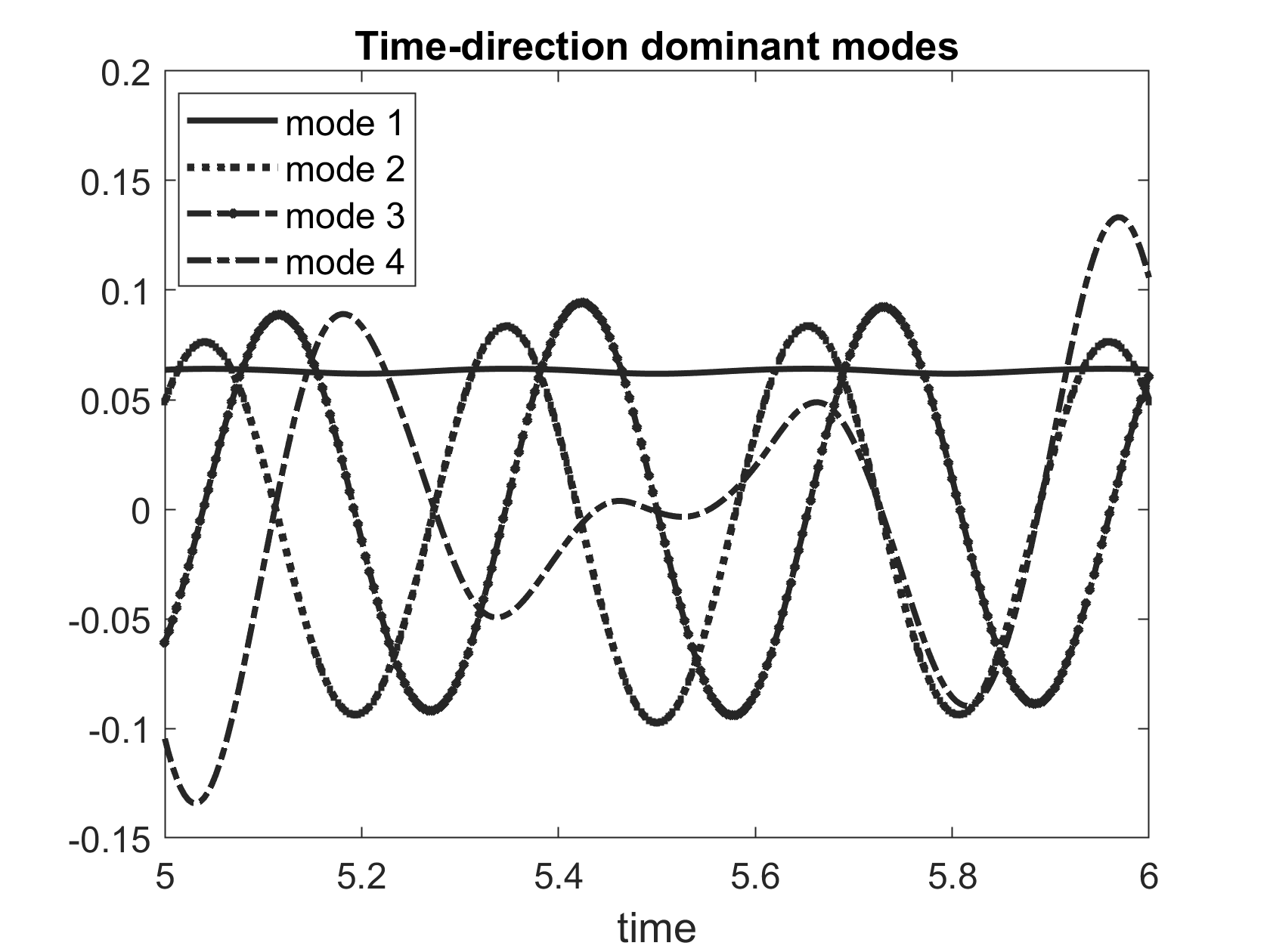}\qquad\includegraphics[width=0.4\textwidth]{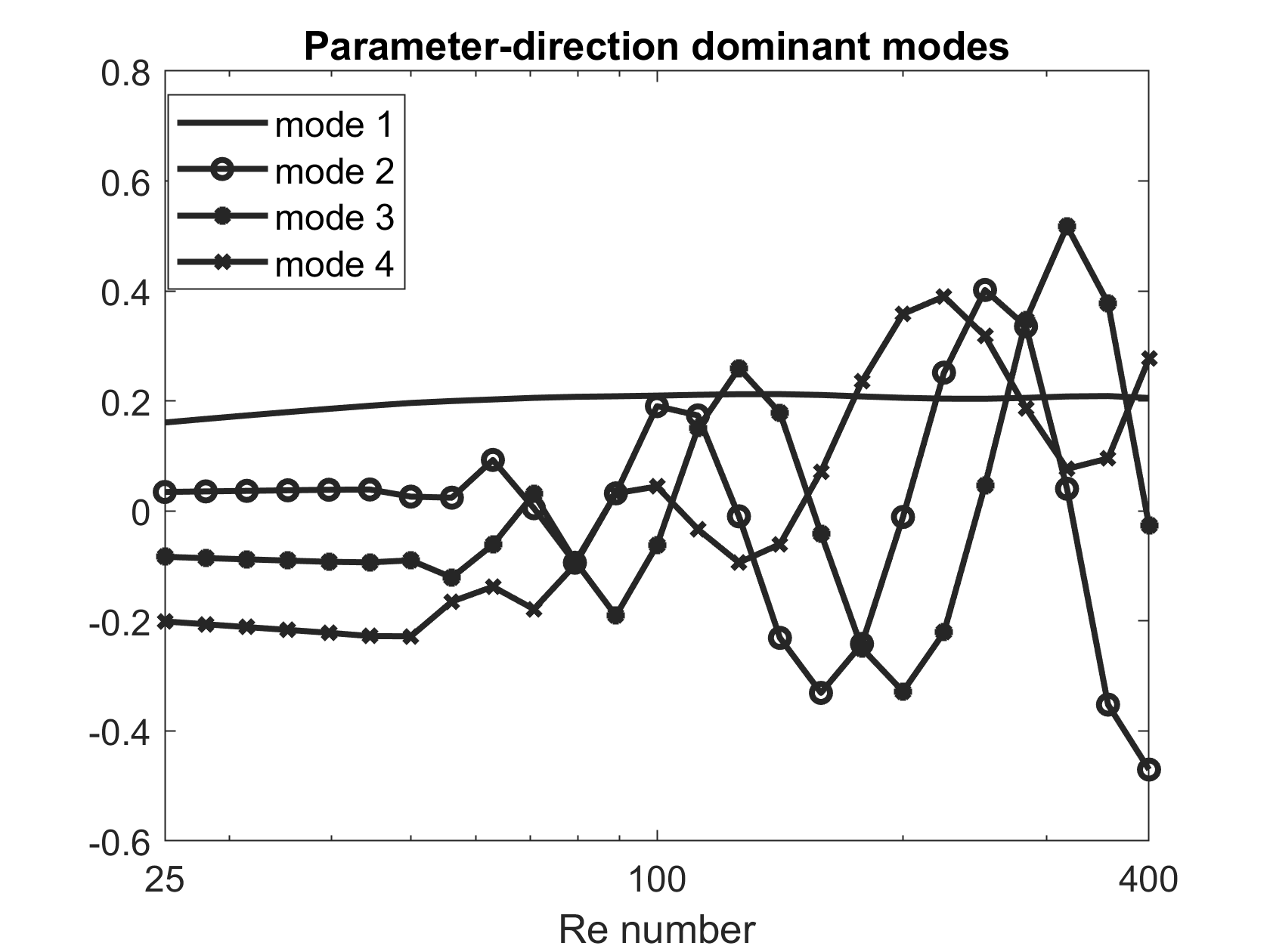}
\end{figure}

Besides building more effective ROM, the LRTD may offer new insights into the properties of  parametric solutions. To give an example, let us consider the  HOSVD singular vectors of $\bPhi$.
Figure~\ref{fig:modes} shows several  dominant vectors in time and parameter directions, which are the first several HOSVD singular vectors of the snapshot tensor in the time and parameter modes. Larger amplitudes of  parameter singular vectors with Re number increase suggest higher sensitivity of flow patterns to the variation of the viscosity   parameter, for  flows with larger Reynolds numbers.

The first singular vectors in time and space direction are approximately constant, cf. Fig.~\ref{fig:modes}.
\begin{wrapfigure}[15]{r}{0.45\textwidth} 
		\includegraphics[width=0.45\textwidth]{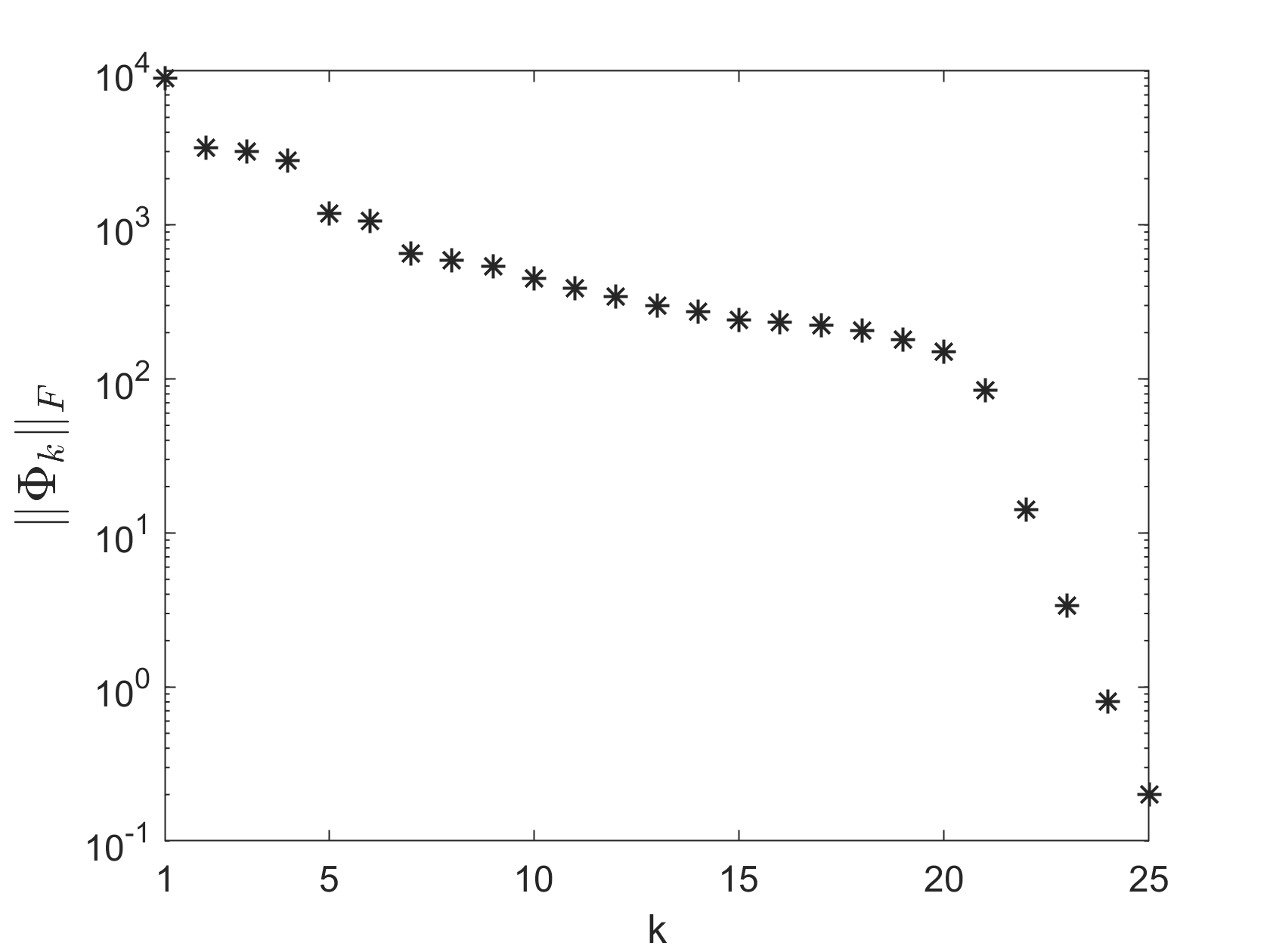}  \vskip-2ex
	\caption{Frobenius norms of space-time stuctures $\Phi_k$ from decomposition \eqref{decomp}.\label{fig:norms}}
\end{wrapfigure}
This indicates that the parametric solution possesses dominant space--parameter and space--time states   which are persistent in time and Reynolds number,  respectively. Let us focus on  persistence in Reynolds number. For  HOSVD, the $\bsigma$ vectors from \eqref{eqn:TDv} are the first $\widetilde K$ singular vectors of the second mode unfolding of $\bPhi$, and so $\bPhi$ can be written as the sum of $K$ direct products:
\begin{equation}\label{decomp}
\bPhi=\sum_{k=1}^K\Phi_k\otimes\bsigma^k,
\end{equation}
where $\Phi_k\in\R^{M\times N}$ are space--time states (note that these are not actual physical states) whose evolution in Reynolds number is determined by $\bsigma^k$. Matrices $\Phi_k$ are mutually orthogonal in the sense of the element-wise product, $\mbox{tr}(\Phi_k\Phi_j^T)=0$ for $k\neq j$, and since $\|\Phi_k\|_F$ equals the $k$-th singular value of the second mode unfolding of $\bPhi$, it also holds that
\[\|\Phi_1\|_F\ge\|\Phi_2\|_F\ge\dots\ge \|\Phi_K\|_F.
\]  
Figure~\ref{fig:norms} shows the norms of the persistent space--time states.  We see that  $\Phi_1$ is not overly dominating and about 20 persistent space--time states contribute to the parametric solution.  
 Using orthogonality of $\bsigma^k$ one finds from \eqref{decomp} that 
\begin{equation}\label{recover} 
	\Phi_k=\bPhi\times_2\bsigma^k.
\end{equation} 
Therefore, $\Phi_k$  are easily recovered for $k=1,\dots,\widetilde K$ once $\bsigma^k$ are provided by HOSVD LRTD.  From \eqref{recover} and the observation that $\bsigma^1$ is nearly constant in Re, we conclude that the dominant space--time state $\Phi_1$ is close to  a scaled average of the  parametric solution in Reynolds number (similar conclusion holds for the dominant space--parameter state --- it is close to a scaled time averaged solution).
 
\begin{figure}[h]	\centering
{\small 1.}\includegraphics[width = .42\textwidth, height=.16\textwidth,viewport=60 20 580 200, clip]{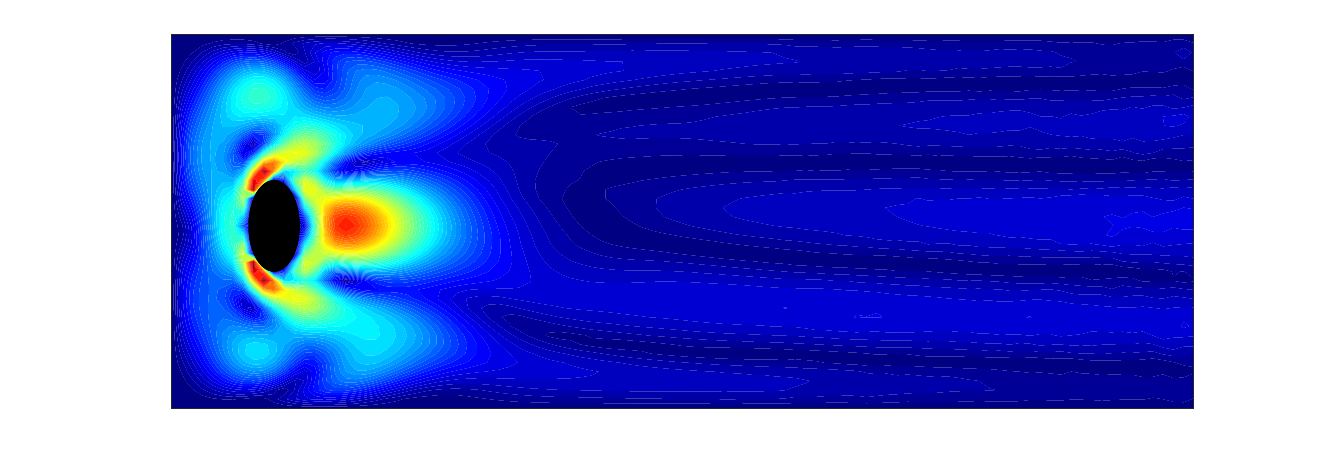}
{\small 2.}\includegraphics[width = .42\textwidth, height=.16\textwidth,viewport=60 20 580 200, clip]{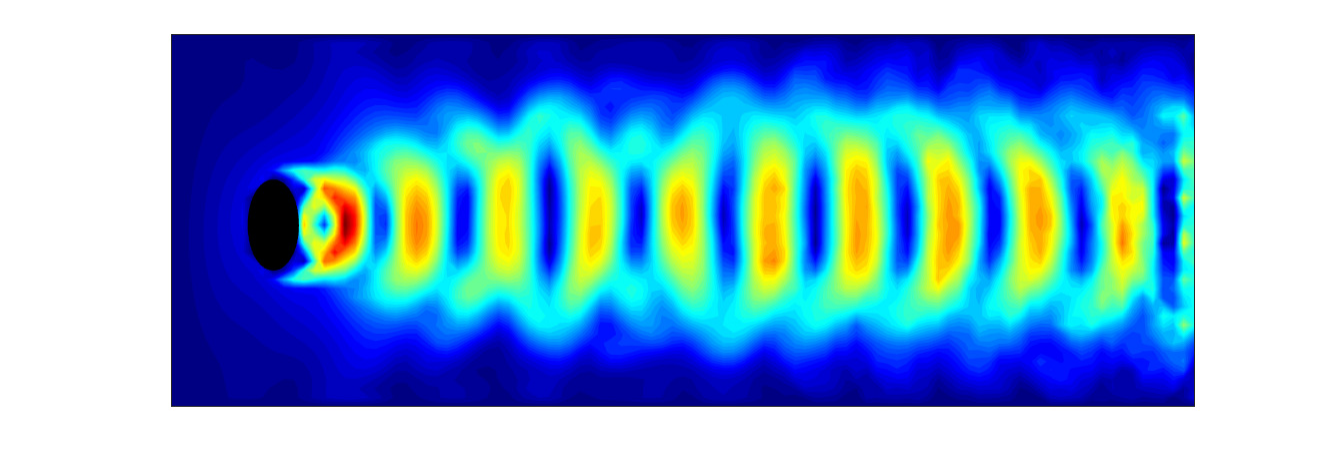}\\
{\small 3.}\includegraphics[width = .42\textwidth, height=.16\textwidth,viewport=60 20 580 200, clip]{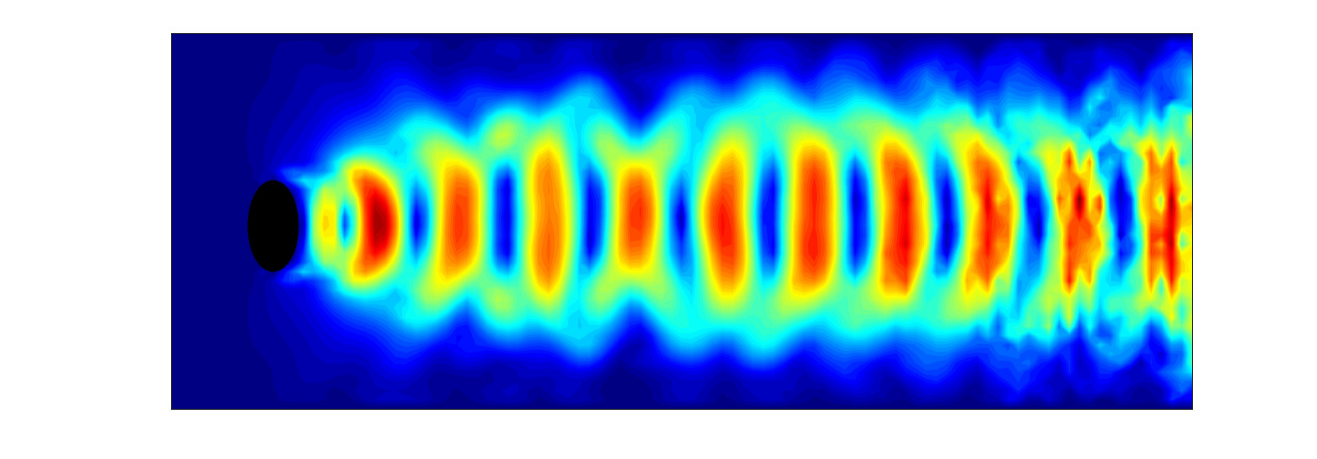}
{\small 4.}\includegraphics[width = .42\textwidth, height=.16\textwidth,viewport=60 20 580 200, clip]{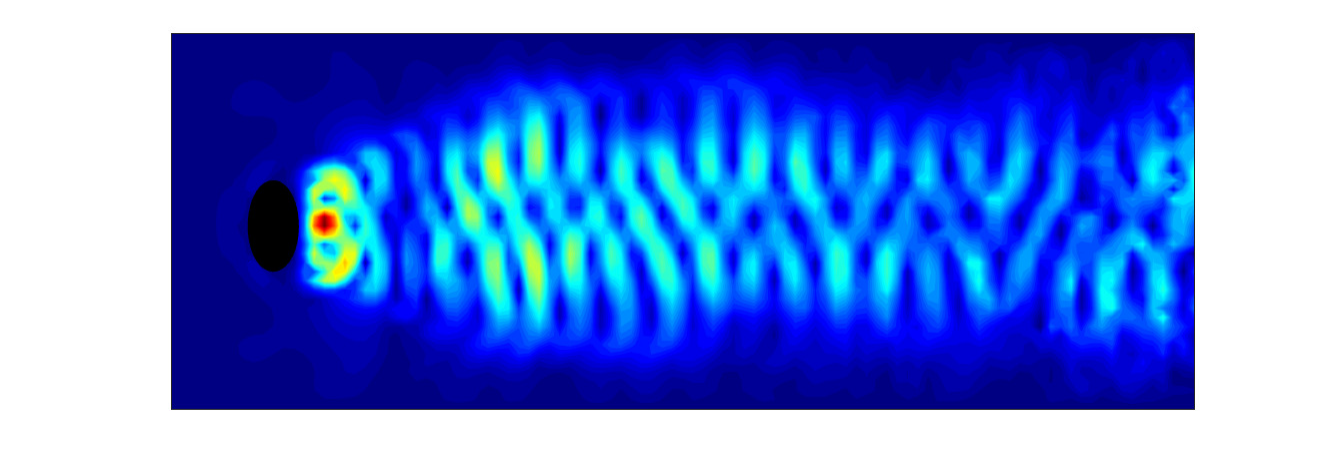}\\
	\caption{\label{modes2} Shown above are the first four spatial modes, taken from $\Phi_1$, the first 
	space-time  persistent state.}
\end{figure}

To gain further insight into the structure of $\Phi_1$, we display in Figure~\ref{modes2} the dominant spatial modes of $\Phi_1$. These are obtained by computing the SVD of $\Phi_1$. Singular values of $\Phi_1$ drop rapidly so that the first spatial mode, shown in Figure~\ref{modes2}, captures nearly 99.9\% of the energy.



\section{Conclusions} The LRTD-ROM, an extension of a POD-ROM for parametric problems, retains essential information about model variation in the parameter domain in a reduced order format. When applied to the incompressible Navier-Stokes equations parameterized with the viscosity coefficient, the LRTD-ROM facilitates accurate prediction of flow statistics along a smooth branch of solutions. Moreover, it enables the identification of parameter structures that may not be apparent through standard POD analysis.
 
 Previously, LRTD-ROMs have demonstrated success in addressing multi-parameter linear and scalar non-linear problems. A natural next step is to extend it to multi-parameter problems of fluid dynamics. Additionally, current research efforts are directed towards developing LRTD-ROMs based on sparse sampling of the parametric domain.   

\section*{Acknowledgments} 
The author M.O. was supported in part by the U.S. National Science Foundation under award DMS-2309197.  The author L.R. was supported by the U.S. National Science Foundation under award DMS-2152623.

This material is based upon work supported by the National Science Foundation under Grant No. DMS-1929284 while the authors were in residence at the Institute for Computational and Experimental Research in Mathematics in Providence, RI, during the semester program.

\bibliographystyle{siamplain}
\bibliography{literatur}{}

\end{document}